\documentclass[pdflatex,sn-mathphys-num]{sn-jnl}

\usepackage{graphicx}%
\usepackage{multirow}%
\usepackage{amsmath,amssymb,amsfonts}%
\usepackage{amsthm}%
\usepackage{mathrsfs}%
\usepackage[title]{appendix}%
\usepackage{xcolor}%
\usepackage{textcomp}%
\usepackage{manyfoot}%
\usepackage{booktabs}%
\usepackage{algorithm}%
\usepackage{algorithmicx}%
\usepackage{algpseudocode}%
\usepackage{listings}%

\theoremstyle{thmstyleone}%
\theoremstyle{thmstyletwo}%

\theoremstyle{thmstylethree}%

\usepackage{lineno}
\usepackage{xcolor}
\newcommand{\response}[1]{\textcolor{black}{#1}}

\begin{document}
\title[Soil texture prediction with Bayesian generalized additive models for spatial compositional data]{Soil texture prediction with Bayesian generalized additive models for spatial compositional data}

\author*[1]{\fnm{Joaquín} \sur{Martínez-Minaya}}\email{jmarmin@eio.upv.es}

\author[2,3]{\fnm{Lore} \sur{Zumeta-Olaskoaga}}\email{lore.zumetaolaskoaga@bio-gipuzkoa.eus}

\author[4]{\fnm{Dae-Jin} \sur{Lee}}\email{daejin.lee@ie.edu}

\affil*[1]{\orgdiv{Department of Applied Statistics and Operational Research and Quality}, \orgname{Universitat Politècnica de València}, \orgaddress{\city{Valencia}, \country{Spain}}}

\affil[2]{\orgname{Biogipuzkoa Health Research Institute}, \orgaddress{\city{Donostia--San Sebastián}, \country{Spain}}}

\affil[3]{\orgdiv{Deusto Physical TherapIker, Faculty of Health Sciences}, \orgname{University of Deusto}, \orgaddress{\city{Donostia--San Sebastián}, \country{Spain}}}

\affil[4]{\orgname{School of Science and Technology, IE University}, \orgaddress{\city{Madrid}, \country{Spain}}}

\abstract{Compositional data (CoDa) play an important role in many fields such as ecology, geology, and biology. The most widely used modeling approaches are based on the Dirichlet and the logistic-normal formulation under Aitchison geometry. Recent developments in simplex geometry allow regression models to be expressed in terms of coordinates and their coefficients to be estimated. Once the model is projected into real space, a multivariate Gaussian regression can be employed. However, most existing methods focus on linear models, and there is a lack of flexible alternatives such as additive or spatial models, especially within a Bayesian framework and with practical implementation details.

In this work, we present a geoadditive regression model for CoDa from a Bayesian perspective using the \texttt{brms} package in R. The model applies the isometric log-ratio (ilr) transformation and penalized splines to incorporate nonlinear effects. We also propose two new Bayesian goodness-of-fit measures for CoDa regression: BR-CoDa-$R^2$ and BM-CoDa-$R^2$, extending the Bayesian $R^2$ to the compositional setting. {These measures, alongside WAIC, support the descriptive assessment of explained variability and complement model evaluation.} The methodology is validated through simulation studies and applied to predict soil texture composition in the Basque Country. Results demonstrate good performance, interpretable spatial patterns, and reliable quantification of explained variability in compositional outcomes.}


\maketitle

\section{Introduction}
\label{sec:Intro}

Understanding the spatial distribution and variability of soil texture is essential for land use planning and other activities related to agricultural management and environmental protection (e.g., prevent soil degradation, preserve soil functions and remediate degraded soil). This work is motivated by the ``Land Use and Cover Area frame Statistical survey'' (LUCAS) project aimed at collecting harmonized data on the state of land use/cover over the extent of the European Union. The study by Ballabio et al. \cite{Ballabio2016} mapped soil properties at a continental scale across Europe. In this work, we aim to map soil texture distribution at a finer scale based on soil samples (at 30 cm depth) surveyed in the Basque Country (Spain) between 2010 and 2018.

\response{Soil texture data are usually treated as compositional data (CoDa), as they are relative percentages of sand, silt and clay, whose sum is constrained to 100\%}. Consequently, analyzing each texture component separately may lead to inconsistent results, such as total values exceeding 100\%. CoDA is the common acronym of Compositional Data Analysis. CoDA theory provides the statistical framework for compositional data, which lies in a simplex sample space \citep{Aitchison1985}. In addition to soil texture samples, we incorporate auxiliary information such as climatic variables (e.g., average precipitation, minimum and maximum temperatures), topographic features (e.g., digital elevation model--DEM), and categorical variables including geological characteristics (e.g., lithology) and land use types (e.g., pastures, extensive crops, vineyards).

Moreover, CoDA is widely applied in other fields such as the analysis of chemical and mineralogical compositions \citep{zhou2018}, economics \citep{Wang2019}, environmental studies of air pollution \citep{aldhurafi2018, Motabertran2022}, research on physical activity and health, and microbiome studies \citep{combettes2020, creus2022bayesian}. Indeed, while the mathematical and statistical foundations of CoDA have been widely developed \citep{pawlowsky2011,pawlowsky2015,pawlowsky2016}, contributions from the statistical modelling perspective remain limited, despite their considerable potential. These include non-parametric density estimation and regression for compositional data \citep{Chacon2010,DiMarzio2015}, model and variable selection \citep{lin2014,susin2020}, model diagnostics \citep{Hijazi2008}, and the development of open-source software tools to support applied researchers working with CoDA \citep{comas2011,lin2014}.

Incorporating geographical information into the model is crucial when the underlying process exhibits spatial dependence. Possible approaches include extending the ordinary kriging to CoDA \citep{Zhang2013}, introducing a conditional autoregressive (CAR) component for discrete spatial domains \citep{Martinez2020}, using a continuous Gaussian field for continuous spaces \citep{martinez-minaya2024, Motabertran2022, pirzamanbein2018}, or employing spatial models with many zeros \citep{Leininger2013, figueira2025unveiling}. In this work, we focus on extending the CoDa regression within the generalized additive models (GAM) framework. GAMs provide a flexible approach to include nonlinear terms, or smooth interactions such as tensor products for spatial anisotropic effects or varying-coefficient terms among other structures \citep{wood2017}. The availability of open-source software R and packages for GAMs such as \texttt{mgcv} facilitates the use of these powerful tools by researchers and practitioners across various fields. Moreover, Bayesian implementations of GAMs and other regression structures (such as multilevel models) are supported by the \texttt{brms} R package, developed by Bürkner \cite{burkner2017}, which provides a high-level R front-end for a wide variety of model types, all fitted using the probabilistic programming language Stan \citep{Betancourt2015,Carpenter2017}.

Model evaluation in CoDA has received increasing attention in recent years. Various approaches have been proposed, including the use of Bayesian model selection criteria such as the Watanabe--Akaike Information Criterion (WAIC; \citealt{gelman2014}), leave-one-out cross-validation (LOOCV), and automatic group construction procedures for leave-group-out cross-validation (LGOCV) adapted for latent Gaussian models fitted via INLA \citep{adin2024automatic, martinez-minaya2024}. These methods support model comparison and out-of-sample validation in compositional frameworks. However, a limitation of these approaches is that they do not provide an interpretable summary of explained variability, analogous to the classical $R^2$.

To fill this gap, we propose two novel Bayesian measures of explained variance in CoDa models, BR-CoDa-$R^2$ and BM-CoDa-$R^2$, which extend the concept introduced by Gelman et al. \cite{gelman2019} to the compositional domain using the ilr transformation. {These measures respect the geometry of the simplex and offer a principled way to summarize explained variability in flexible regression settings.} In this regard, we also show how CoDa regression can naturally be extended to nonlinear and spatial frameworks through Bayesian GAMs.

The paper is structured as follows. In \autoref{sec:SoilTextureBasqueCountry}, we provide an overview of the soil texture survey conducted in the Basque Country. In \autoref{sec:Simplicial}, we introduce the fundamental concepts of the simplex space underlying CoDa and its associated geometry. In \autoref{sec:GAMsCoDA}, we present the foundations of simplicial regression and generalized additive models (GAMs), and describe how these structures are defined within a Bayesian framework. In \autoref{sec:BayesianModelSelectionCoDA}, we introduce several Bayesian model evaluation criteria, which lead to the presentation of our proposed Bayesian CoDa-$R^2$ measures in \autoref{sec:BayesianRsquared}. \autoref{sec:Simulation} \response{presents a series of illustrative examples that evaluate the performance of the proposed methodology and the goodness-of-fit measures introduced}, followed by a real-data application in \autoref{sec:CoDAGAMSoil}. Finally, \autoref{sec:Conclusions} summarizes the main findings and outlines potential avenues for future research.


\section{Soil texture survey in the Basque Country}
\label{sec:SoilTextureBasqueCountry}

Soil texture is one of the most fundamental soil characteristics. It influences various physical, chemical, biological, and hydrological properties and processes, such as water and nutrient retention and infiltration. {It is commonly determined by sedimentation methods using a pipette or a hydrometer, which provide continuous estimates of the relative proportions of soil particle-size fractions: sand whose particle diameter varies between $2.00$ and $0.05\,\text{mm}$; silt, which varies between $0.05$ and $0.002\,\text{mm}$; and clay, whose diameter is lower than $0.002\,\text{mm}$. Importantly, these proportions are not obtained by counting individual particles, but are derived from sedimentation-based measurement procedures. As a result, soil texture data consist of continuous proportions that sum to a constant and contain only relative information, rather than discrete counts. For this reason, they are naturally treated as CoDa}.

In recent years, numerous studies have proposed different approaches to map soil texture. Notable examples include Adhikari et al. \cite{adhikari2013} and Ballabio et al. \cite{Ballabio2016}. Among other methods, Akpa et al. \cite{akpa2014} and Chagas et al. \cite{chagas2016} use random forest and multiple linear regression; Niang et al. \cite{niang2014}, Gooley et al. \cite{gooley2014} or Hengl et al. \cite{Hengl2014} employ techniques such as kriging and Support Vector Machine (SVM); Wadoux and Molnar \cite{Wadoux2021} conduct analysis with interpretable machine learning methods or Swetha et al. \cite{swetha2020} with convolutional neural network (CNN) algorithms; and authors such as Buchanan et al. \cite{Buchanan2012} who compare ordinary kriging with geoadditive models, Poggio and Gimona \cite{poggio2017} who use hybrid geostatistical GAMs or Nussbaum et al. \cite{Nussbaum2017} who employ boosted geoadditive models. In most cases, covariates used in the models were derived from the DEM and remote sensing.

The spatial distribution of soil texture plays a key role in many of these approaches, as it can help address major global environmental challenges such as climate change, food and water scarcity, and biodiversity loss \cite{hartemink2008}. However, the relevance of spatial effects is not limited to this context. In many CoDa problems, incorporating spatial components is crucial for improving predictive performance. Spatial CoDa problems can be defined over discrete domains, where autoregressive spatial effects are required \cite{Martinez2020}, or over continuous domains, where classical kriging methods are typically employed \cite{pawlowsky2016,Zhang2013}. From a Bayesian perspective, several authors have also addressed this problem \cite{pirzamanbein2018,martinez-minaya2024}. GAMs offer a suitable approach for integrating geographic information. They enable the inclusion of effects such as elevation in a nonlinear fashion, providing greater flexibility to explain the phenomena of interest \cite{Hastie1990}. Thus, the combination of GAMs and CoDa provides an ideal methodological framework for mapping soil texture in our study area, the Basque Country (Spain).

\begin{figure}
\centering
\includegraphics[width=\textwidth]{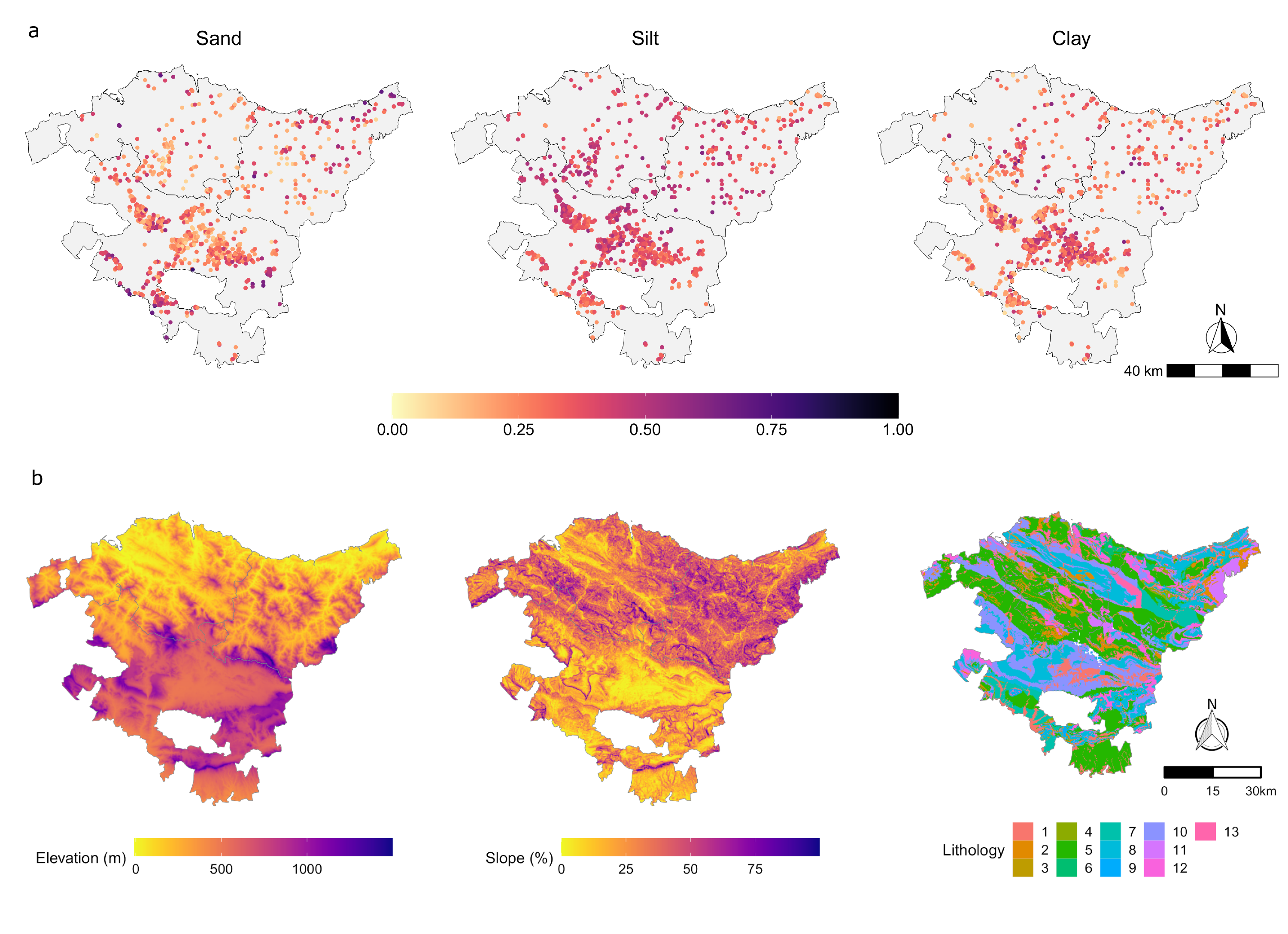}
\caption{(a) Map of the study area showing the relative abundances of sand, silt, and clay at each sampling location. (b) Map of the study area showing the three covariates used in the analysis: elevation, slope, and lithology.}
\label{fig:data_soil}
\end{figure}

The available data consist of 2279 soil samples (at $30\,\text{cm}$ depth) surveyed in the Basque Country between 2010--2018 (Figure~\ref{fig:data_soil}), work carried out by the Basque Institute for Agricultural Research and Development (\url{https://neiker.eus/en/}). Variables such as Elevation, Slope, and Lithology were obtained from \url{https://www.geo.euskadi.eus/}. The spatial resolution of the covariates was $500\,\text{m} \times 500\,\text{m}$. See \autoref{fig:data_soil} for a graphical representation of the study area and covariates, and \autoref{table:litho} for a detailed description of the Lithology factor levels.

\begin{table}
\caption{Categorical levels of the Lithology covariate used in the modeling framework.}
\label{table:litho}
\centering
\begin{tabular}{r l}
\toprule
Factor level & Description \\
\midrule
1 & Surface Deposits \\
2 & Coarse-grained detrital rocks (sandstones). Dominant \\
3 & Medium-grained detrital rocks (siltstones). Dominant \\
4 & Fine-grained detrital rocks (shales). Dominant \\
5 & Alternating detrital clasts \\
6 & Decarbonate marls \\
7 & Marls \\
8 & Impure limestones and calcarenites \\
9 & Clays with gypsum and other salts \\
10 & Alternation of marl limestones, chalky marls and calcarenites \\
11 & Slates \\
12 & Limestones and dolomites \\
13 & Volcanic rocks \\
\bottomrule
\end{tabular}
\end{table}

\section{The simplex and its geometry} 
\label{sec:Simplicial}

CoDA relies on understanding the mathematical structure of the simplex, the appropriate sample space in which compositions reside. Unlike traditional Euclidean space, the simplex imposes specific constraints, such as the constant sum of the components, which require specialized geometric tools to properly describe and analyze the data. In this section, we introduce the essential concepts of Aitchison geometry, which allow us to define operations such as distance, orthogonality, and inner products for CoDa. We then describe the isometric log-ratio (ilr) transformation, a key tool that allows mapping data from the simplex to an unconstrained Euclidean space, thereby enabling the application of classical statistical techniques. Lastly, we explore statistical summaries within the simplicial framework, including the definitions of center and variance in the context of CoDA, providing the foundation for the modeling approaches that follow.

\subsection{Aitchison Geometry}

A composition is a $D$-component vector whose components show the relative importance of $D$ parts forming a total. It is characterized by the fact that only relative information is relevant, scaling it by any constant values does not change its meaning \cite{aitchison1982}. The simplex, $S^D$, is the sample space of CoDa, and it is commonly defined as:
\begin{equation}
S^D = \left\{ \boldsymbol{y} = [y_1, \ldots, y_D]; \ y_d > 0,\ d = 1, \ldots, D;\ \sum_{d=1}^D y_d = \kappa \right\},
\end{equation}
being $\kappa$ an arbitrary positive constant, e.g., $\kappa = 1$ if measurements are performed in parts per unit (proportions) or $\kappa = 100$ if they are in percentages. From now on, and without loss of generality, we assume that $\kappa = 1$. As in the ordinary real Euclidean space, a specific geometry is defined in the simplex, also known as simplicial geometry. It does not follow ordinary Euclidean rules observed in real vector space. Nevertheless, concepts such as distance between points, straight lines, orthogonality, norm, and inner vector product exist for a simplex vector space \cite{egozcue2011}. 

Perturbation, denoted by $\oplus$ (an internal operation), and powering, denoted by $\odot$ (an external operation), are defined as the two main operations in the simplex. They correspond to addition and scalar multiplication in real space. Thus, if $\boldsymbol{z}, \boldsymbol{y} \in S^D$, perturbation in the simplex is defined as
\begin{equation}
\boldsymbol{z} \oplus \boldsymbol{y} = \frac{[z_1y_1, \ldots, z_Dy_D]}{z_1y_1 + \ldots + z_Dy_D} = C(z_1y_1, \ldots, z_Dy_D),
\end{equation}
where $C$ denotes the closure operator which standardizes the vector dividing each component by the sum of its components. Similarly, for any real number $\alpha$, the powering operation is defined as
\begin{equation}
\alpha \odot \boldsymbol{y} = C(y_1^{\alpha}, \ldots, y_D^{\alpha}).
\end{equation}

Moreover, an inner product $\langle \cdot , \cdot \rangle_a$ can be established as
\begin{equation}
\langle \boldsymbol{z}, \boldsymbol{y} \rangle_a = \sum_{d=1}^D \log \frac{z_d}{g_m(\boldsymbol{z})} \log \frac{y_d}{g_m(\boldsymbol{y})},
\label{eq:inner_aitch}
\end{equation}
where $g_m(\boldsymbol{z})$ and $g_m(\boldsymbol{y})$ denote the geometric mean of the components of $\boldsymbol{z}$ and $\boldsymbol{y}$ respectively. With all these operations, $(S^D, \oplus, \odot, \langle \cdot , \cdot \rangle_a)$ has a $(D-1)$-dimensional real Euclidean vector space structure, and the geometry defined on the simplex is named simplicial or Aitchison geometry \cite{pawlowsky2001}.

The related norm $\|\cdot\|_a$ and distance $d_a(\cdot, \cdot)$ are defined as
\begin{equation}
\|\boldsymbol{y}\|_a = \langle \boldsymbol{y}, \boldsymbol{y} \rangle_a^{1/2},
\end{equation}
and
\begin{equation}
d_a(\boldsymbol{z}, \boldsymbol{y}) = \|\boldsymbol{z} \ominus \boldsymbol{y}\|_a,
\end{equation}
being $\ominus$ the subtraction operation defined as $\boldsymbol{z} \ominus \boldsymbol{y} = \boldsymbol{z} \oplus ((-1)\odot \boldsymbol{y})$.

\subsection{The isometric log-ratio transformation (ilr)}

Due to the Euclidean structure of $S^D$, an orthonormal basis with respect to the inner product can be constructed \cite{Egozcue2003}, and a composition $\boldsymbol{y} \in S^D$ can be represented by its coordinates with respect to such a basis:
\begin{equation}
\boldsymbol{y} = \bigoplus_{d=1}^{D-1} y_d^{*} \odot \boldsymbol{e}_d ,\quad y_d^{*} = \langle \boldsymbol{y}, \boldsymbol{e}_d \rangle_a ,
\label{eq:coord_basis}
\end{equation}
where $\{\boldsymbol{e}_1, \ldots, \boldsymbol{e}_{D-1}\}$ is an orthonormal basis of $S^D$, and $\boldsymbol{y}^{*} = [y_1^{*}, \ldots, y_{D-1}^{*}]$ denotes the coordinate vector of $\boldsymbol{y}$ with respect to the selected basis. The function ilr: $S^D \rightarrow \mathbb{R}^{D-1}$, which assigns the coordinates $\boldsymbol{y}^{*}$ to $\boldsymbol{y}$, i.e. ilr$(\boldsymbol{y}) = \boldsymbol{y}^{*}$, is called isometric log-ratio transformation, and constitutes an isometric isomorphism between vector spaces \cite{pawlowsky2015}.

As ilr$(\boldsymbol{y}) \in \mathbb{R}^{D-1}$ can be identified with the coordinates of $\boldsymbol{y} \in S^D$ with respect to the basis $\{\boldsymbol{e}_1, \ldots, \boldsymbol{e}_{D-1}\}$, it can also be identified with coordinates with respect to the canonical basis in $\mathbb{R}^{D-1}$, $\{\boldsymbol{e}_1^{*}, \ldots, \boldsymbol{e}_{D-1}^{*}\}$. This brings us to the equivalent definition of ilr coordinates: ilr$(\boldsymbol{y}) = \sum_{d=1}^{D-1} \langle \boldsymbol{y}, \boldsymbol{e}_d \rangle_a \boldsymbol{e}_d^{*}$, where ilr$(\boldsymbol{e}_d) = \boldsymbol{e}_d^{*}$, $d = 1, \ldots, D-1$. As pointed out in Egozcue et al. \cite{Egozcue2003}, the inverse ilr transformation corresponds to the reconstruction of $\boldsymbol{y}$ in the reference basis of $S^D$.
\begin{equation}
\boldsymbol{y} = \text{ilr}^{-1}(\boldsymbol{y}^{*}) = \bigoplus_{d=1}^{D-1} (\langle \boldsymbol{y}^{*}, \boldsymbol{e}_d^{*} \rangle \odot \boldsymbol{e}_d).
\end{equation}
{Different choices of orthonormal bases in $S^D$ lead to alternative ilr representations, which are related through orthogonal transformations in $\mathbb{R}^{D-1}$. Consequently, the methodology developed in this work is invariant to the particular choice of ilr transformation.}

{Among the various ways to define orthonormal bases in $S^D$, those linked to a sequential binary partition (SBP) are particularly popular due to their interpretability \cite{egozcue2005,pawlowsky2011}. Their main appeal lies in the fact that they allow compositions to be expressed in terms of grouped parts. The Cartesian coordinates of a composition in such a basis are known as balances. \response{For concreteness, we adopt the ilr coordinates as computed by default in the R package \texttt{compositions} \citep{gerald2021,boogaart2013}, originally proposed by \citet{Egozcue2003}. The corresponding basis elements $e_d$, $d = 1, \ldots, D-1$, are given by}
\begin{equation}
e_d = C \left[
\exp \left(
\underbrace{-\sqrt{\frac{1}{d(d+1)}}, \ldots, -\sqrt{\frac{1}{d(d+1)}}}_{d \text{ elements}},
\sqrt{\frac{d}{d+1}}, 0, \ldots, 0
\right)
\right],
\label{eq:basis}
\end{equation}
\response{which form an orthonormal basis of $S^D$ with respect to the Aitchison inner product (\autoref{eq:inner_aitch}). Under this choice, the ilr coordinates $\boldsymbol{y}^{*} = \mathrm{ilr}(\boldsymbol{y})$ are given by}
\begin{equation}
y_d^{*} = -\sqrt{\frac{d}{d+1}} \log \left( \frac{g_m(y_1, \ldots, y_d)}{y_{d+1}} \right),\quad d = 1,2, \ldots, D-1 .
\label{eq:ilr}
\end{equation}
{Besides the ilr transformation, other log-ratio representations such as the additive log-ratio (alr) and centered log-ratio (clr) are also commonly used \cite{Aitchison1985}. We focus on ilr coordinates due to their isometric properties, which ensure a faithful representation of the simplex geometry in Euclidean space.}

\subsection{Moments of simplicial statistics}

When addressing random compositions within a simplex, it is necessary to define certain key concepts. These include, among others, the identification of the center and the variance of a random composition. As shown in Egozcue and Pawlowsky-Glahn \cite{egozcue2011}, the center of a random composition $\boldsymbol{y}$ can be defined as:
\begin{equation}
\text{Cen}(\boldsymbol{y}) = \text{ilr}^{-1}\operatorname{E}(\text{ilr}(\boldsymbol{y})) = C (\exp(\operatorname{E}(\log(\boldsymbol{y})))) \,,
\end{equation}
where $\operatorname{E}(\cdot)$ denotes the ordinary expectation in the ordinary real Euclidean space. Regarding the second order moments, a common metric for total variance of a random composition is defined as
\begin{equation}
\operatorname{Var}_{\mathrm{Tot}}(\boldsymbol{y}) = \operatorname{E}(d_a^2(\boldsymbol{y}, \text{Cen}(\boldsymbol{y}))) \,.
\label{eq:total_variance}
\end{equation}
However, this is not the only way to express the total variance of a random composition. It can also be reformulated as the variance decomposition of the ilr coordinates, i.e.,
\begin{equation}
\operatorname{Var}_{\mathrm{Tot}}(\boldsymbol{y}) = \sum_{d = 1}^{D-1} \operatorname{Var}(y_d^*) \,,
\label{eq:totalvar}
\end{equation}
where $y_d^*$ denotes the $d$th-coordinate of the random composition $\boldsymbol{y}$. As proved in Egozcue and Pawlowsky-Glahn \cite{egozcue2005}, the decomposition holds when the Aitchison distance is expressed in terms of orthonormal coordinates. The second-order structure is completed by the covariances between coordinates $\operatorname{Cov}(y_j^*, y_d^*)$ arranged in the $(j,d)$-entry of the $(D-1)\times(D-1)$ covariance matrix $\boldsymbol{\Sigma}$ associated with the ilr coordinates. {An important consequence of this representation is that $\operatorname{Var}_{\mathrm{Tot}}(\boldsymbol{y})$ does not depend on the particular ilr basis used. Indeed, if $\boldsymbol{y}^*$ and $\tilde{\boldsymbol{y}}^*$ denote two ilr representations of the same random composition $\boldsymbol{y}$ associated with two orthonormal bases of $S^D$, then there exists an orthogonal matrix $\mathbf{Q}$ such that $\tilde{\boldsymbol{y}}^* = \mathbf{Q}\boldsymbol{y}^*$. Therefore, their covariance matrices satisfy $\tilde{\boldsymbol{\Sigma}} = \mathbf{Q}\boldsymbol{\Sigma}\mathbf{Q}^\top$. Since
\[
\operatorname{Var}_{\mathrm{Tot}}(\boldsymbol{y})=\sum_{d=1}^{D-1}\operatorname{Var}(y_d^*)=\operatorname{tr}(\boldsymbol{\Sigma}),
\]
and the trace is invariant under orthogonal similarity transformations, it follows that $\operatorname{Var}_{\mathrm{Tot}}(\boldsymbol{y})$ is invariant to the choice of ilr coordinates. Consequently, any global measure built from sums of variances in orthonormal ilr coordinates is also invariant to the selected ilr basis.}


\section{Bayesian Spatial Generalized Additive Models for CoDA}
\label{sec:GAMsCoDA}

\subsection{The simplicial regression}

Any kind of regression is primarily evaluated by its residuals. In the literature, in the context of CoDa, different assumptions have been proposed for modeling residuals. The most popular are the assumption of Dirichlet residuals or logistic-normal residuals. In this work, we focus on the logistic-normal distribution introduced by Aitchison and Shen \cite{Aitchison1980}. They {show} that if $\boldsymbol{y} \in S^D$, whose representation in coordinates with respect to a selected orthonormal basis is $\boldsymbol{y}^* \in \mathbb{R}^{D-1}$, $\boldsymbol{y}^* = \mathrm{ilr}(\boldsymbol{y})$, the random composition $\boldsymbol{y}$ has a logistic-normal distribution or, equivalently, a normal distribution in the simplex, whenever $\boldsymbol{y}^*$ has a multivariate normal distribution, i.e., $\boldsymbol{y}^* \sim \mathcal{N}(\boldsymbol{\mu}, \boldsymbol{\Sigma})$.

{Once the ilr transformation has been applied, the compositional constraint that characterizes CoDa is removed, allowing the use of standard multivariate Gaussian regression techniques. Each transformed observation is represented as a $(D-1)$-dimensional vector. Let $\boldsymbol{y}_1^*, \ldots, \boldsymbol{y}_N^*$ denote the ilr-transformed observations corresponding to $\boldsymbol{y}_1, \ldots, \boldsymbol{y}_N \in S^D$. The regression model is then defined as}
\begin{equation}
\begin{aligned}
\boldsymbol{y}_n^* &\sim \mathcal{N}\left(\boldsymbol{\mu}_n, \boldsymbol{\Sigma}\right), \\
\mu_{dn} &= \eta_{dn}, \quad d = 1,\ldots,D-1,\; n = 1,\ldots,N,
\end{aligned}
\label{eq:mu}
\end{equation}
{where $\boldsymbol{\mu}_n$ is the $(D-1)$-dimensional mean vector for observation $n$, $\mu_{dn}$ denotes its $d$th coordinate}, $\boldsymbol{\Sigma}$ is the variance-covariance matrix formed by the variances in the diagonal $\Sigma_{dd} = \sigma_d^2$, $d = 1,\ldots,D-1$, and covariances $\Sigma_{dk} = \rho_{kd}\sigma_k\sigma_d$, for $k,d = 1,\ldots,D-1$, and $k \neq d$. $\eta_{dn}$ denotes the linear predictor associated with the $d$th coordinate of the $n$th observation. {Since} we are in the ordinary real Euclidean space, we deal with a multivariate Gaussian likelihood, which, as a member of the exponential family, allows the linear predictor to include different terms such as linear terms, spatial components, splines, etc. Finally, the inverse ilr transformation is applied to reconstruct compositions in $S^D$ from the fitted coordinates \cite{pawlowsky2016}. In the next section, we provide the multivariate model for extending the linear predictor within a GAM framework.

\subsection{Bayesian Generalized Additive models}

{GAMs are flexible models that allow for the incorporation of linear and nonlinear effects within a regression framework for a response distribution in the exponential family \cite{Hastie1990}. Let $Y_1, \ldots, Y_N$ denote the response variables and $y_1, \ldots, y_N$ their observed values. A common formulation of the model is given by}
\begin{equation}
\eta_n = g(\mu_n) = \boldsymbol{X}_n \boldsymbol{\beta} + \sum_{j=1}^J f_j(x_{jn}), \quad n = 1, \ldots, N,
\label{eq:gam}
\end{equation}
{where $\mu_n = \operatorname{E}(Y_n)$, $g(\cdot)$ is a known link function, $\boldsymbol{X}_n \in \mathbb{R}^p$ is the $n$th row of the design matrix for the parametric component, $\boldsymbol{\beta} \in \mathbb{R}^p$ is the corresponding vector of unknown coefficients, and $f_j$ are unknown smooth functions of the covariates $x_j$, typically univariate, although multivariate smooths are also possible.}

{Bivariate smoothing is particularly relevant in our setting as soil texture exhibits strong spatial structure, with nearby locations tending to have similar compositions due to underlying geological and environmental processes. This induces spatial dependence that cannot be fully explained by the observed covariates alone. Within the GAM framework, this is represented through smooth functions of the form \( f_j(x_1, x_2) \), which can be interpreted as spatial surfaces capturing residual spatial variability. In our model, one of these smooth terms is explicitly used to represent a spatial effect based on geographical coordinates. To estimate the unknown smooth functions, GAMs commonly rely on basis expansions. In particular, each smooth term can be written as}
\[
f_j(x)=\sum_{k=1}^{K_j} b_{jk}(x)\theta_{jk}
= \boldsymbol{B}_j(x)^\top \boldsymbol{\theta}_j,
\]
{where \(b_{jk}(\cdot)\) are known basis functions and \(\boldsymbol{\theta}_j\) is the vector of spline coefficients. In this work, we construct these smooth terms using P-splines. P-splines are a specific formulation that combines a B-spline basis with a discrete difference penalty on the coefficients. This provides a flexible yet computationally efficient way to control smoothness. Compared to other spline-based approaches, such as thin plate splines, P-splines offer a convenient balance between flexibility, interpretability, and computational tractability, particularly in settings involving large datasets or multiple smooth components \cite{eilers2021}.} \response{Under this representation, estimation of GAMs is typically based on penalized regression. For a generic predictor, the spline coefficients are obtained by maximizing a penalized log-likelihood of the form}
\[
\ell(\boldsymbol{\beta},\boldsymbol{\theta}_j)-\frac{1}{2}\sum_{j=1}^J \lambda_j
\boldsymbol{\theta}_j^\top \mathbf{S}_j \boldsymbol{\theta}_j,
\]
\response{where \(\ell(\cdot)\) denotes the log-likelihood, \(\mathbf{S}_j\) is the penalty matrix associated with the \(j\)th smooth term, and \(\lambda_j\) is a smoothing parameter controlling the trade-off between fidelity to the data and smoothness of the fitted function. For P-splines, \(\mathbf{S}_j\) is usually induced by first or second order differences between adjacent spline coefficients, so that rough coefficient sequences are penalized more heavily than smooth ones.}

\response{From a Bayesian perspective, this penalized log-likelihood corresponds to placing a Gaussian prior directly on the full vector of B-spline coefficients,}
\[
\boldsymbol{\theta}_j \sim \mathcal{N}\!\left(\boldsymbol{0},\,\sigma_j^2\mathbf{S}_j^{-}\right),
\]
\response{where \(\mathbf{S}_j^{-}\) denotes a generalized inverse of the (rank-deficient) penalty matrix. For a second-order P-spline penalty, this is the classical Bayesian P-spline prior of Lang and Brezger \cite{langbrezger2004}, under which the coefficients \(\theta_{jk}\) follow a second-order random walk (RW2): consecutive coefficients are correlated through \(\mathbf{S}_j^{-}\), rather than independent.}

\response{A key fact, particularly relevant in this paper, is that this formulation admits an equivalent mixed model representation \cite{wood2017,linzhang1999,fahrmeir2004,eilers2015}. Let \(\mathbf{S}_j = \mathbf{U}_j\boldsymbol{\Lambda}_j\mathbf{U}_j^\top\) denote the eigendecomposition of the penalty matrix. Its null space, of dimension equal to the order of the difference penalty (2 for a second-order penalty), is collected in a matrix \(\mathbf{X}_j^F\), while the eigenvectors associated with the nonzero eigenvalues, rescaled by the inverse square root of their eigenvalues, are collected in \(\mathbf{X}_j^R\). This provides an orthogonal reparametrization of the coefficient vector,}
\[
\boldsymbol{\theta}_j = \mathbf{X}_j^F\boldsymbol{\beta}_j^F + \mathbf{X}_j^R\boldsymbol{u}_j,
\]
\response{where, to avoid confusion with the original B-spline weights \(\boldsymbol{\theta}_j\), we denote by \(\boldsymbol{\beta}_j^F\) the coefficients of the unpenalized component and by \(\boldsymbol{u}_j\) the reparametrized penalized coefficients. Under this representation, \(\boldsymbol{u}_j\) can be treated as i.i.d.\ Gaussian random effects,}
\[
\boldsymbol{u}_j \sim \mathcal{N}\!\left(\boldsymbol{0},\,\sigma_j^2\mathbf{I}\right),
\]
\response{while \(\boldsymbol{\beta}_j^F\) is assigned a fixed-effect-type prior, as for the other linear terms in the model. Under this formulation, the smoothing parameter \(\lambda_j\) is directly related to the variance component \(\sigma_j^2\): larger penalties correspond to smaller prior variances and therefore smoother functions, whereas smaller penalties allow more flexible fits.}

\response{In this work, all P-spline terms are constructed using a second-order difference penalty, so that the null space of \(\mathbf{S}_j\) has dimension $2$, spanned by the constant and linear polynomial terms. Posterior inference for the resulting model is carried out via Hamiltonian Monte Carlo (specifically, the No-U-Turn sampler) as implemented in Stan, through the interface provided by \texttt{brms} \cite{burkner2017}; internally, \texttt{brms} exploits the mixed model reparametrization described above, fitting the model directly in terms of \(\boldsymbol{\beta}_j^F\) and \(\mathbf{u}_j\) rather than the original \(\boldsymbol{\theta}_j\). The fixed-effect component \(\boldsymbol{\beta}_j^F\) is assigned a prior as any other fixed effect in the model, while the penalized component \(\mathbf{u}_j\) is assigned the proper Gaussian prior derived above. Its variance component \(\sigma_j^2\) is treated as an unknown parameter and assigned a prior, specifically, we use \texttt{brms}'s default half-Student-\(t\) prior with three degrees of freedom, \(\sigma_j\sim\text{Half-}t(3,0,s)\), with scale \(s\) determined internally from the data, so that the amount of smoothing for each term is learned directly from the data rather than imposed a priori.} 

\response{The same reparametrization extends to the bivariate, tensor-product P-spline used to represent the spatial effect. Combining the null and range spaces of the two marginal bases yields one unpenalized (bilinear) component, spanning the joint null space of both marginals, and three separately penalized components, each associated with its own variance parameter, following the general tensor-product mixed-model construction of \citet{woodScheiplFaraway2013} and the ANOVA-type decomposition of \citet{leeDurban2011}.} Finally, we can include a GAM in the context of CoDA. After applying the ilr transformation, each coordinate is modeled through its own linear predictor of the form given in \autoref{eq:gam}: 
\begin{align}
\begin{split}
\boldsymbol{y}_n^* &\sim \mathcal{N}(\boldsymbol{\mu}_n, \boldsymbol{\Sigma}) \\
\mu_{dn} &= \boldsymbol{X}_n \boldsymbol{\beta}^d + \sum_{j=1}^{J^d} f^d_j(x_{jn})\,,
\end{split}
\label{eq:general_CoDA_reg}
\end{align}
{for \(d=1,\ldots,D-1\) and \(n=1,\ldots,N\). Therefore, each ilr coordinate can contain its own set of linear, nonlinear, and spatial effects, while the covariance matrix \(\boldsymbol{\Sigma}\) captures the dependence between coordinates.}

\section{{Bayesian model evaluation: goodness-of-fit and predictive criteria}}
\label{sec:BayesianModelSelectionCoDA}

This section is devoted to introduce some measures to compare Bayesian additive models. In particular, we focus on WAIC \cite{gelman2014} and the Bayesian $R^2$ \cite{gelman2019}.

\subsection{Watanabe-Akaike information criterion (WAIC)}

WAIC is a fully Bayesian approach which has proven to be a reliable approximation for estimating the out-of-sample expectation. It is based on the computation of the log pointwise posterior predictive density (lppd), along with a correction for the effective number of parameters ($p_{\text{WAIC}}$) to adjust for model complexity \cite{gelman2014,Vehtari2017}. WAIC is defined as
\begin{equation}
\text{WAIC} = -2\,lppd + 2p_{\text{WAIC}},
\end{equation}
being 
\[
lppd = \sum_{n=1}^N \log \left( \frac{1}{S} \sum_{s=1}^S p(y_n \mid \theta^s) \right),
\quad
p_{\text{WAIC}} = \sum_{n=1}^N V_{s=1}^S \big( \log(p(y_n \mid \theta^s)) \big).
\]
A lower WAIC indicates a better model fit, adjusted for model complexity. Note that we follow the notation from Gelman et al. \cite{gelman2019}, where $V$ denotes the sample variance of any vector $\boldsymbol{z} = (z_1,\ldots,z_J)$, defined as
\[
V_{j=1}^J z_j = \frac{1}{J-1} \sum_{j=1}^J (z_j - \bar{z})^2,
\quad \text{with} \quad
\bar{z} = \frac{1}{J} \sum_{j=1}^J z_j.
\]

\subsection{Bayesian $R^2$}

The coefficient of determination, $R^2$, is a widely used measure of goodness of fit that quantifies the proportion of variability in the response explained by the predictors. Classically, it is defined as the ratio between the variance of the predicted values and the total variance of the observed data. However, in Bayesian models, this definition presents a drawback: the variance of the predicted values can sometimes exceed the total variance of the data, which can result in $R^2 > 1$. Gelman et al. \cite{gelman2019} addressed this issue by proposing a Bayesian $R^2$ that is well-defined based on the distribution of future data $(y_n)$ and the expected values conditional on the unknown parameters $y_n^{\text{pred}} = \operatorname{E}(y_n \mid X_n, \theta)$. Here, $X_n$ denotes the set of predictors associated with $y_n$. For a linear model, $y_n^{\text{pred}}$ corresponds to the linear predictor, while for a generalized linear model, it is the linear predictor transformed to the data scale. The Bayesian $R^2$ is then defined as
\begin{equation}
R^2 = \frac{\text{Explained variance}}{\text{Explained variance} + \text{Residual variance}} = \frac{\mathrm{var}_{fit}}{\mathrm{var}_{fit} + \mathrm{var}_{res}},
\end{equation}
where $\mathrm{var}_{fit} = V_{n=1}^N \operatorname{E}(y_n \mid \theta) = V_{n=1}^N y_n^{\text{pred}}$ represents the variance of the predictive means, and $\mathrm{var}_{res} = \operatorname{E}\big( V_{n=1}^N (y_n - y_n^{\text{pred}}) \mid \theta \big)$ is the modeled residual variance. Both $\mathrm{var}_{fit}$ and $\mathrm{var}_{res}$ are defined conditionally on the model parameters $\theta$, therefore the resulting $R^2$ is conditional on $\theta$. As discussed in Gelman et al. \cite{gelman2019}, extending the definition of $R^2$ to the Bayesian context simply requires acknowledging that inference is based on posterior simulation draws $\theta^s$, with $s = 1,\ldots,S$. For each posterior draw $\theta^s$, the $R^2$ can be computed as
\begin{equation}
R^2_s = \frac{V_{n=1}^N y_n^{\text{pred},s}}{V_{n=1}^N y_n^{\text{pred},s} + \mathrm{var}_{res}^s},
\end{equation}
where $y_n^{\text{pred},s} = \operatorname{E}(y_n \mid X_n, \theta^s)$ denotes the predictive mean for draw $s$, and $\mathrm{var}_{res}^s$ is the corresponding residual variance for that draw. As discussed in Vehtari et al. \cite{vehtariBayesR2}, there are two main approaches for defining the modeled residual variance in the Bayesian $R^2$. The first is the \emph{residual-based} approach, which uses draws from the residual distribution and computes
\begin{equation}
\mathrm{var}_{res}^s = V_{n=1}^N e_n^s,
\end{equation}
where $e_n^s = y_n - y_n^{\text{pred},s}$.

The second is the \emph{model-based} approach, which relies on draws from the modeled residual variances. For example, in linear regression,
\begin{equation}
\mathrm{var}_{res}^s = (\sigma^2)^s,
\end{equation}
while in logistic regression,
\begin{equation}
\mathrm{var}_{res}^s = \frac{1}{N} \sum_{n=1}^N \pi_n^s (1 - \pi_n^s).
\end{equation}
While in this work we follow the posterior-based formulation of $R^2$ proposed by Gelman et al. \cite{gelman2019} and Vehtari et al. \cite{vehtariBayesR2}, alternative approaches have emerged that leverage the same definition of $R^2$ from a different perspective. For instance, Zhang et al. \cite{zhang2020} and Aguilar and Bürkner \cite{aguilar2023} propose placing prior distributions directly on the $R^2$ coefficient as a means to construct informative priors on regression coefficients. These methods are primarily designed for prior specification and model regularization in high-dimensional or hierarchical settings, rather than as goodness-of-fit metrics. Although these perspectives are valuable in their own context, our focus here remains on the use of $R^2$ as a posterior predictive measure for model evaluation.

\section{An extension of Bayesian $R^2$ for CoDa}
\label{sec:BayesianRsquared}

We extend the concept of $R^2$ to the CoDa context by grounding it in the total variability of a composition expressed through the variance decomposition of ilr coordinates (see \autoref{eq:totalvar}). This formulation, proposed by Egozcue and Pawlowsky-Glahn \cite{egozcue2011}, establishes the total variance as the sum of variances of the orthonormal ilr coordinates, thus fundamentally respecting the Aitchison geometry of the simplex. While this framework provides the core foundation for our CoDa-$R^2$, related ideas were already explored by Hijazi and Jernigan \cite{Hijazi2009}, who proposed alternative $R^2$ measures grounded in compositional variance. Our proposal builds directly on the ilr-based decomposition, offering a natural extension to the Bayesian setting. Accordingly, the explained variability of CoDa can be expressed as the total variance of the predicted ilr coordinates:
\begin{equation}
\mathrm{var}_{fit} = \sum_{d=1}^{D-1} V_{n=1}^N \left( y_{dn}^{*,\mathrm{pred}} \right),
\end{equation}
where $y_{dn}^{*,\mathrm{pred}}$ denotes the expected value of the $d$th ilr coordinate for the $n$th observation, conditional on the covariates $X_n$ and the model parameters:
\begin{equation}
y_{dn}^{*,\mathrm{pred}} = \operatorname{E}\left( y_{dn}^* \mid X_n, \theta \right).
\end{equation}
{Therefore, this construction is invariant to the particular ilr basis used, since it is ultimately based on sums of variances of orthonormal coordinates, or equivalently, traces of covariance matrices.}

As in the univariate case described by Gelman et al. \cite{gelman2019}, bringing this definition to the Bayesian context involves computing the CoDa-$R^2$ for each posterior draw $\theta^s$, with $s = 1,\ldots,S$. For each draw, the CoDa-$R^2$ is given by
\begin{equation}
\mathrm{CoDa}\text{-}R^2_s =
\frac{\sum_{d=1}^{D-1} V_{n=1}^N \left( y_{dn}^{*,\mathrm{pred},s} \right)}
{\sum_{d=1}^{D-1} V_{n=1}^N \left( y_{dn}^{*,\mathrm{pred},s} \right) + \mathrm{var}_{res}^s},
\end{equation}
where $y_{dn}^{*,\mathrm{pred},s} = \operatorname{E}\left( y_{dn}^* \mid X_n, \theta^s \right)$ represents the predictive mean of the $d$th ilr coordinate for observation $n$ at posterior draw $s$. In line with Gelman et al. \cite{gelman2019} and Vehtari et al. \cite{vehtariBayesR2}, we consider two alternative formulations for $\mathrm{var}_{res}^s$:

\begin{enumerate}
\item \emph{Residual-based CoDa-$R^2$ (BR-CoDa-$R^2$)}:
\begin{equation}
\mathrm{var}_{res}^s =
\sum_{d=1}^{D-1} V_{n=1}^N \left( y_{dn}^{*} - y_{dn}^{*,\mathrm{pred},s} \right),
\end{equation}
where $y_{dn}^{*}$ is the observed ilr coordinate.

\item \emph{Model-based CoDa-$R^2$ (BM-CoDa-$R^2$)}:
\begin{equation}
\mathrm{var}_{res}^s =
\sum_{d=1}^{D-1} \Sigma_{dd}^s,
\end{equation}
where $\Sigma_{dd}^s$ denotes the posterior draw of the variance of the $d$th ilr coordinate.
\end{enumerate}
Both formulations provide coherent, geometry-respecting measures of explained variability for CoDa models, fully compatible with the Bayesian paradigm. As a direct consequence of the invariance properties of sums of variances in orthonormal ilr coordinates, the resulting CoDa-$R^2$ measure is invariant to the particular ilr basis used. In practice, minor differences may arise due to Monte Carlo variability, but these are negligible and consistent with the theoretical invariance of the measure. As with any Bayesian goodness-of-fit measure, \response{the proposed CoDa-$R^2$ is computed from posterior quantities and therefore inherits any sensitivity to prior specifications from the fitted model.}

{In what follows, we demonstrate how these measures can be used to summarize and compare the explained variability across models. From an applied perspective, they provide an interpretable summary of explanatory capacity in compositional settings, complementing predictive criteria such as WAIC in environmental and spatial applications.} CoDa-$R^2$ enables a principled quantification of the probability that one model explains more variability than another. The Bayesian framework naturally provides posterior distributions for CoDa-$R^2$, making it straightforward to compute probabilities such as $P\left(\mathrm{CoDa}\text{-}R^2_{\mathrm{M1}} \geq \mathrm{CoDa}\text{-}R^2_{\mathrm{M2}}\right)$ when comparing two competing models, M1 and M2. By fixing the credible level at $\alpha = 0.1$, we consider a substantial difference if this probability exceeds $1 - \alpha = 0.9$, indicating that M1 explains substantially more variability than M2. If the probability is below $\alpha = 0.1$, we conclude that M2 explains substantially more variability than M1. Probabilities lying between $0.1$ and $0.9$ suggest that both models have a similar explanatory capacity. This threshold serves as a pragmatic decision rule rather than a formal hypothesis test, and the choice of $\alpha$ corresponds to the level of uncertainty we are willing to tolerate in this comparison. 

{It is important to note that CoDa-$R^2$ should be interpreted as an in-sample summary of explained variability rather than a standalone criterion for model selection.} {In addition, the posterior distribution of CoDa-$R^2$ provides a direct way to quantify the uncertainty associated with the explained variability. This allows practitioners to assess how precisely the model captures the variability in the data, and to evaluate whether the level of explained variability is sufficiently reliable for the intended application, rather than relying solely on a single point estimate. This tool is particularly useful when comparing nested models, as it allows us to quantify how the inclusion of a new variable changes the explained variability when moving from a reduced model to an extended one. However, when covariates are correlated, such changes should be interpreted with caution, since they depend on the variables already included in the model and therefore on the order in which they are added. In this sense, CoDa-$R^2$ is a useful and interpretable summary of explanatory capacity in applications involving CoDa, but not an order-invariant decomposition of variable importance.}

\section{Simulation examples}
\label{sec:Simulation}
\response{This section presents three illustrative examples, designed to evaluate the performance of the proposed methodology under controlled, known data-generating mechanisms.} The first example assesses the ability of a Bayesian linear CoDa model to recover the true parameters and hyperparameters. The second example evaluates the accuracy of CoDa regression when smooth terms are incorporated into the linear predictor, including both univariate and bivariate smooth components. The final example compares BM-CoDa-$R^2$ and BR-CoDa-$R^2$ for model discrimination under varying data-generating mechanisms. \response{The purpose of these examples is to demonstrate that the proposed Bayesian CoDa-GAM framework can be fitted in practice using existing software (\texttt{brms}), and to verify that it recovers known ground truth under controlled settings, rather than to characterize its operating properties across replicates, which lies beyond the scope of this illustration.}

\subsection{Bayesian CoDa linear regression}

The first simulation considers CoDa in $S^4$. A sample of 100 realizations was drawn from a 3-dimensional Gaussian distribution with the following structure:
\begin{equation}
\label{eq:struct_linear_regression}
\begin{split}
\boldsymbol{y}^*_{n} &\sim \mathcal{N}(\boldsymbol{\mu}_{n}, \boldsymbol{\Sigma}) \\
\mu_{dn} &= \beta_0^d + \beta_1^d x_{1n} + \beta_2^d x_{2n} + \beta_3^d x_{3n}.
\end{split}
\end{equation}
where $x_{kn}$, $k = 1,\ldots,3$, $n = 1,\ldots,100$, are covariates drawn independently from a Uniform$(0,1)$ distribution, and $\beta_k^d$ are the fixed parameters corresponding to the $d$th ilr coordinate and the $k$th covariate, with $k = 0$ denoting the intercept. The intercepts were fixed to $\beta_0^1 = 1$, $\beta_0^2 = -0.5$, and $\beta_0^3 = -2$. For the slopes, we specified $\beta_1^1 = -0.5$, $\beta_2^1 = 1$, $\beta_3^1 = -0.5$ for the first ilr coordinate; $\beta_1^2 = 1$, $\beta_2^2 = -1$, $\beta_3^2 = 0$ for the second; and $\beta_1^3 = 1$, $\beta_2^3 = -0.5$, $\beta_3^3 = -0.5$ for the third. The standard deviations were fixed at $\sigma_1 = 0.10$, $\sigma_2 = 0.05$, and $\sigma_3 = 0.08$, with correlation parameters $\rho_{12} = 0.5$, $\rho_{13} = 0.2$, and $\rho_{23} = 0.8$. These small standard deviations reduce noise, facilitating the assessment of parameter recovery. After generating the data in $\mathbb{R}^3$, the inverse ilr transformation was applied to obtain simulated compositional data in $S^4$.

\response{Bayesian inference was conducted by assigning prior distributions to both regression coefficients and covariance parameters. Specifically, we used zero-centered Gaussian priors with standard deviation 10 for the fixed effects, in place of the improper flat prior that \texttt{brms} assigns to this class by default; this choice provides weak regularization on the scale of the linear predictor, allowing the data to dominate the inference while avoiding extreme coefficient values. Given the scale of the covariates and the magnitude of the true regression parameters considered in this simulation, this prior is sufficiently diffuse to ensure that posterior estimates are primarily driven by the likelihood. For the standard deviations and the correlation matrix, we retained \texttt{brms}'s own default priors \cite{burkner2017}: a half Student-$t$ distribution with 3 degrees of freedom for the standard deviations, which provides regularization while allowing for moderate variability; and a Lewandowski, Kurowicka and Joe prior (LKJ) \cite{Lewandowski2009}, a distribution defined over correlation matrices that allows controlling the degree of shrinkage towards independence, with the regularization parameter left at its default value of 1, which corresponds to a uniform prior over all valid correlation matrices \cite{burkner2017}.}

To demonstrate that the model is capable of recovering the true parameter values, \autoref{table:posterior_linear_regression} reports the posterior means and 95\% credible intervals for the fixed effects, standard deviations, and correlations. In all cases, the true values lie within the corresponding 95\% credible intervals. Model fit was further assessed by comparing posterior predictive distributions with the original data. This visualization allows for a direct comparison between the predicted compositions and the true values, providing an intuitive validation of the model fit. Further details can be found in \autoref{fig:fitted_linear_regression} (Appendix).

\begin{table}[h]
\caption{Means and 95\% credible intervals for the sample posterior distributions. Comparison with real values.}
\label{table:posterior_linear_regression}%
\centering
\begin{tabular}{@{}lrrrrr@{}}
\toprule
\multicolumn{6}{c}{Fixed effects} \\
\midrule
ilr coordinate & Parameter & Real value & Mean & Q 2.5\% & Q 97.5\% \\
\midrule
First 
& $\beta_0$ & 1.000  & 0.896 & 0.449 & 1.370 \\
& $\beta_1$ & -0.500 & -0.479 & -0.501 & -0.457 \\
& $\beta_2$ & 1.000  & 0.993 & 0.965 & 1.023 \\
& $\beta_3$ & -0.500 & -0.506 & -0.527 & -0.484 \\[4pt]
Second
& $\beta_0$ & -0.500 & -0.543 & -0.730 & -0.357 \\
& $\beta_1$ & 1.000  & 0.998 & 0.989 & 1.007 \\
& $\beta_2$ & -1.000 & -1.000 & -1.012 & -0.989 \\
& $\beta_3$ & 0.000  & 0.006 & -0.003 & 0.015 \\[4pt]
Third
& $\beta_0$ & -2.000 & -1.981 & -2.278 & -1.676 \\
& $\beta_1$ & 1.000  & 0.993 & 0.979 & 1.006 \\
& $\beta_2$ & -0.500 & -0.504 & -0.523 & -0.486 \\
& $\beta_3$ & -0.500 & -0.491 & -0.504 & -0.477 \\
\midrule
\multicolumn{6}{c}{Hyperparameters} \\
\midrule
& Parameter & Real value & Mean & Q 2.5\% & Q 97.5\% \\
\midrule
& $\sigma_1$  & 0.100 & 0.108 & 0.094 & 0.125 \\
& $\sigma_2$  & 0.050 & 0.045 & 0.039 & 0.051 \\
& $\sigma_3$  & 0.080 & 0.070 & 0.061 & 0.081 \\
& $\rho_{12}$ & 0.500 & 0.554 & 0.403 & 0.679 \\
& $\rho_{13}$ & 0.200 & 0.252 & 0.061 & 0.426 \\
& $\rho_{23}$ & 0.800 & 0.786 & 0.699 & 0.853 \\
\botrule
\end{tabular}
\end{table}

\subsection{Bayesian CoDa GAM}

This second example considers CoDa in $S^3$. A total of 100 realizations were generated from a bivariate Gaussian distribution with the following structure:
\begin{equation}
\label{eq:struct_spline}
\begin{split}
\boldsymbol{y}^*_{n} &\sim \mathcal{N}\left( \boldsymbol{\mu}_{n}, \boldsymbol{\Sigma} \right), \\
\mu_{dn} &= f_1^d(xs_{1n}) + f_2^d(xs_{2n}, xs_{3n}),
\end{split}
\end{equation}
where $f_1^d(xs_{1n})$ denotes a univariate smooth function for the $d$th ilr coordinate and covariate $xs_{1n}$ in the $n$th observation, and $f_2^d(xs_{2n}, xs_{3n})$ denotes a bivariate smooth function for the $d$th ilr coordinate and covariates $xs_{2n}$ and $xs_{3n}$. The covariates $xs_{kn}$, with $k = 1,\ldots,3$, were simulated from a Uniform$(0,1)$ distribution. The functions were defined as
\begin{equation}
\label{eq:GuWahba}
\begin{split}
f_1^1(xs_1) &= 0.2\,xs_1^{11}(10(1-xs_1))^6 \\
&\quad + 10(10xs_1)^3(1-xs_1)^{10},
\end{split}
\end{equation}
\begin{equation}
f_1^2(xs_1) = \sin(2\pi xs_1).
\end{equation}
It is worth noting that \autoref{eq:GuWahba} corresponds to one of the components of the classic four-term additive model introduced by Gu and Wahba \cite{guwahba1991}, widely used as a benchmark for testing smooth term recovery. It was simulated using the function \texttt{gamSim} from the R package \texttt{mgcv} \cite{wood2017}. The same R package was used to simulate two bivariate smooth functions:
\begin{equation}
\label{eq:f2_1}
\begin{split}
f_2^1(xs_2, xs_3) &= (\pi^{0.3}0.4)\Big(1.2\exp\big(-\frac{(xs_2-0.2)^2}{0.3^2} - \frac{(xs_3-0.3)^2}{0.4^2}\big) \\
&\quad + 0.8\exp\big(-\frac{(xs_2-0.7)^2}{0.3^2} - \frac{(xs_3-0.8)^2}{0.4^2}\big)\Big),
\end{split}
\end{equation}
\begin{equation}
\label{eq:f2_2}
f_2^2(xs_2, xs_3) = (\pi^{0.3}0.4)\,1.2\exp\left(-\frac{(xs_2-0.5)^2}{0.3^2} - \frac{(xs_3-0.5)^2}{0.4^2}\right).
\end{equation}

All functions were subsequently centered. The standard deviations and correlation parameter were set to 0.03, 0.05, and 0.5, respectively. Finally, applying the inverse ilr transformation allowed us to map the data to $S^3$ (see \autoref{fig::simplex_tetra_sim_spl}). The main objective of this example is to reconstruct the nonlinear functions.

\begin{figure}[H]
    \centering
    \includegraphics[width=0.7\textwidth]{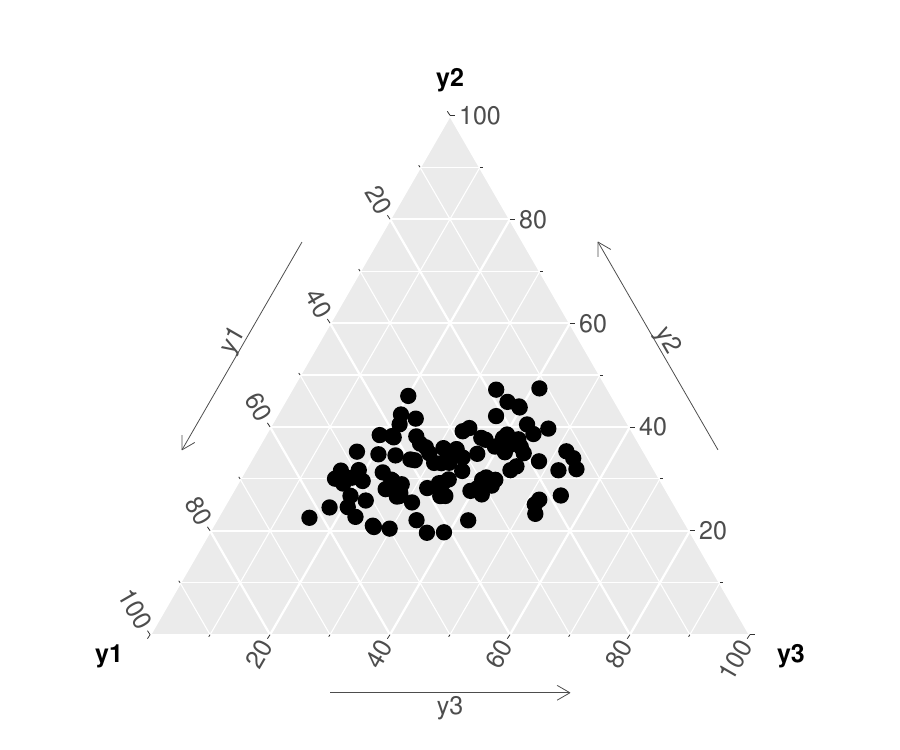}
    \caption{Simulated data in $S^3$. The composition is represented in percentages.}
    \label{fig::simplex_tetra_sim_spl}
\end{figure}

For inference, we adopted the same functional structure as in the data-generating mechanism, replacing $f_1^d(\cdot)$ with $s^d(\cdot)$ to represent univariate penalized P-splines, and $f_2^d(\cdot)$ with $t^d(\cdot)$ to denote tensor-product smooths based on P-spline bases. \response{Both were specified with basis dimensions of $k=5$ and $k=10$, compared in the sensitivity analysis below}.

\response{Gaussian priors with mean zero and standard deviation 10 were specified for the fixed-effect coefficients, which include the unpenalized (linear) component \(\boldsymbol{\beta}_j^F\) of each P-spline term (see \autoref{sec:GAMsCoDA}). The variance components \(\sigma_j^2\) governing the penalized part \(\mathbf{u}_j\) of each smooth term were assigned \texttt{brms}'s default half-Student-\(t\) prior with three degrees of freedom (\autoref{sec:GAMsCoDA}). The same default prior was used for the remaining standard deviations of the model, and the correlation matrix was assigned a Lewandowski, Kurowicka and Joe (LKJ) prior with regularization parameter set to 1.}

The fitted model successfully recovers the underlying smooth effects, as illustrated in \autoref{fig:s_sim_splines} and \autoref{fig:t_sim_splines}. Additionally, \autoref{fig:variances_sim_splines} shows the posterior distributions of the standard deviations and correlation parameter, confirming that the true values are well captured by the model. To further assess the performance of the model, we also represented the mean posterior predictive distribution alongside the original data. This visualization illustrates how well the model recovers the true underlying compositional structure in the simplex after applying the inverse ilr transformation. The plot provided in \autoref{fig:fitted_sim_splines} (Appendix) offers a clear diagnostic of the predictive fit, confirming the model's ability to accurately capture the complex nonlinear relationships present in the simulated data.

\begin{figure}[H]
    \centering
    \includegraphics[width=\textwidth]{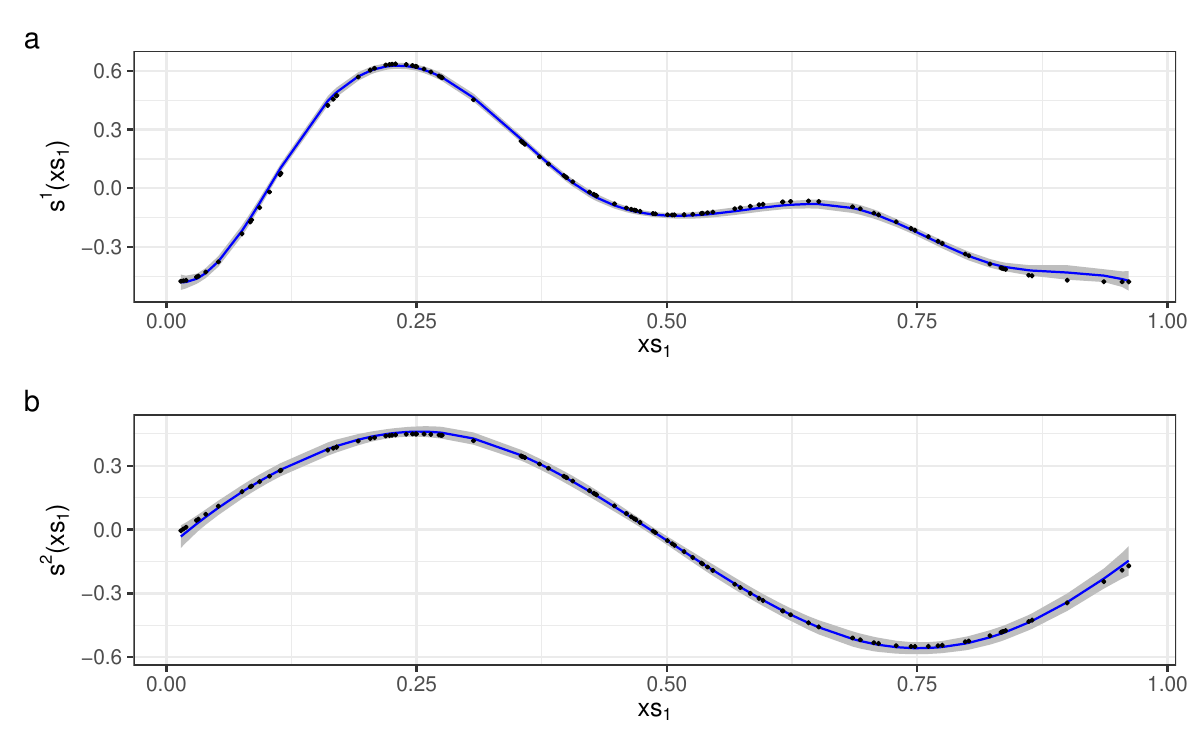}
    \caption{Points represent the simulated data from the one-dimensional smooth functions $f_1^d$, with $d = 1, 2$. The blue lines and gray regions correspond to the posterior mean and the 95\% credible intervals, respectively, for the estimated smooth terms $s^d(xs_1)$, with $d = 1, 2$.}
    \label{fig:s_sim_splines}
\end{figure}

\begin{figure}[h]
    \centering
    \includegraphics[width=\textwidth]{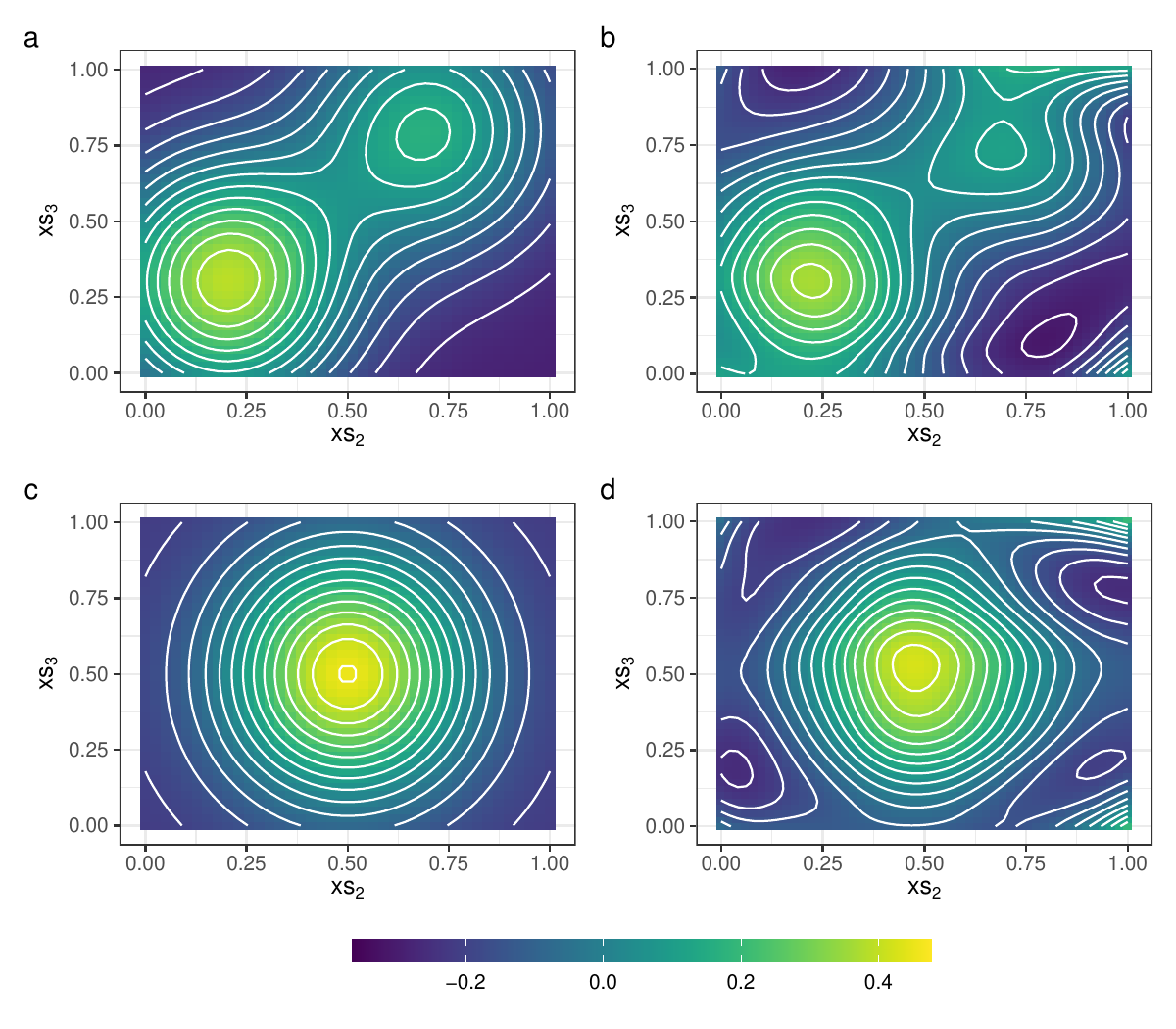}
    \caption{(a) and (c) Contour plots of the simulated bivariate smooth functions $f_2^1$ and $f_2^2$. (b) and (d) Posterior mean estimates of the smooth terms $t^1(x_{s2}, x_{s3})$ and $t^2(x_{s2}, x_{s3})$, respectively.}
    \label{fig:t_sim_splines}
\end{figure}

\begin{figure}[h]
    \centering
    \includegraphics[width=\textwidth]{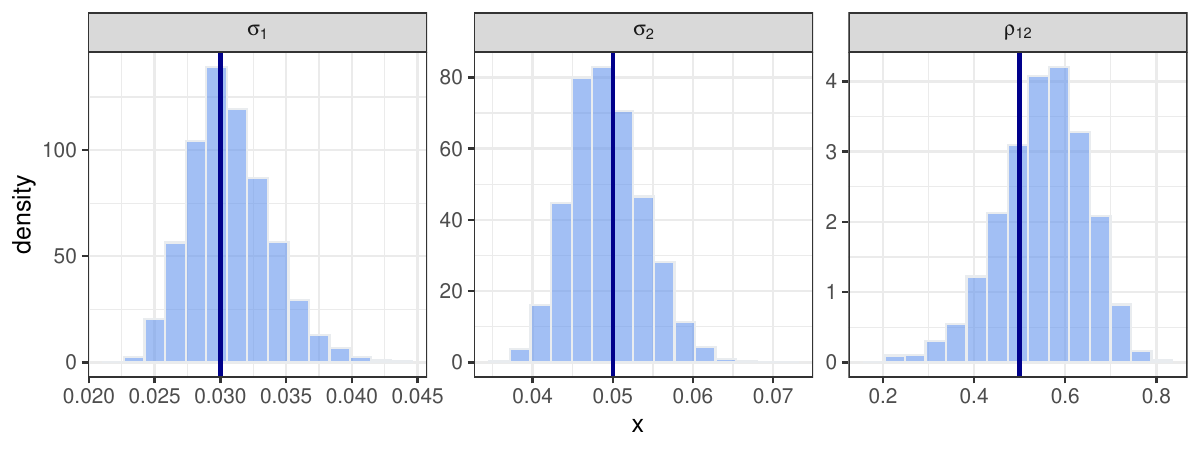}
    \caption{Histograms with the posterior distribution for the standard deviations $\sigma_1$ and $\sigma_2$, and the correlation parameter $\rho_{12}$. Vertical lines correspond with the real values from the simulation.}
    \label{fig:variances_sim_splines}
\end{figure}

\response{We also examined how these results hold up under noisier or more strongly correlated data-generating settings, repeating the simulation across a full factorial grid of noise multipliers ($\kappa \in \{1,2,3,5,7,10\}$, scaling $\boldsymbol{\Sigma}$ in \autoref{eq:struct_spline}) and ilr-coordinate correlations ($\rho \in \{0.1, 0.5, 0.9\}$), for $6 \times 3 = 18$ combinations in total. Recovery of the univariate and bivariate smooth terms remained accurate across all noise levels and correlation values, with some expected degradation of the tensor-product surfaces at the highest noise levels; the corresponding CoDa-$R^2$ measures decreased monotonically with $\kappa$, consistently across all three correlation levels. Full results are provided in Online Resource 1 (Figures~S1--S7). Finally, to assess sensitivity to the prior specification, we refitted this same GAM simulation under several alternative prior choices for the spline coefficients and variance components. Resulting CoDa-$R^2$ values changed negligibly across specifications, suggesting robustness of the measure under reasonable prior choices (see Online Resource 2 for details).}

\subsection{{Comparison of Explained Variability Using CoDa-$R^2$}}
Using the simulated data and model structures described in \autoref{eq:struct_linear_regression} and \autoref{eq:struct_spline}, we fitted several models and computed WAIC along with our proposed measures: BR-CoDa-$R^2$ and BM-CoDa-$R^2$. In the linear regression case, we generated an additional covariate, $x_4$, which was not involved in the data-generating process. This allowed us to evaluate the behavior of our proposed measures in the presence of an irrelevant covariate. {In the GAM case, no additional covariates were included; instead, we assessed model flexibility by varying the number of spline basis functions ($k = 5$ and $k = 10$). Increasing $k$ results in greater model flexibility and more complex spline shapes. {
In this context, varying the number of basis functions ($k$) allows us to assess the impact of the basis dimension on the model’s ability to capture nonlinear patterns. We observe that smaller values of $k$ may limit flexibility, while larger values provide a better approximation of more complex structures. This is consistent with the general principle in GAMs that the basis dimension should be chosen sufficiently large to avoid underfitting, with the effective complexity of the model controlled by the smoothing penalty \cite{wood2017}. CoDa-$R^2$ is used here as an in-sample summary of explained variability and is interpreted alongside WAIC, which provides a predictive criterion that accounts for model complexity and out-of-sample predictive performance.}}

\autoref{table:models_sim} summarizes the predictors used in each model for both the linear regression and GAM settings, along with the corresponding WAIC values and posterior means (and standard deviations) of BR-CoDa-$R^2$ and BM-CoDa-$R^2$.
\begin{table}[h]
\caption{{Linear predictors for the different fitted models. WAIC and the posterior mean (standard deviation) of BR-CoDa-$R^2$ and BM-CoDa-$R^2$ are reported to summarize predictive fit and explained variability, respectively.}}
\label{table:models_sim}%
\centering
\begin{tabular}{@{}llrrr@{}}
\toprule
\multicolumn{5}{c}{Linear regression} \\
\midrule
Code & Model & $WAIC$ & BR-CoDa-$R^2$ & BM-CoDa-$R^2$ \\
\midrule
M1 & $\beta^d_0 + \beta^d_1 x_{1n} + \beta^d_2 x_{2n} + \beta^d_3 x_{3n}$ & -884.760 & 0.996 (0.001) & 0.996 (0.001) \\
M2 & $\beta^d_0 + \beta^d_1 x_{1n} + \beta^d_2 x_{2n} + \beta^d_4 x_{4n}$ & -324.312 & 0.890 (0.003) & 0.888 (0.010) \\
M3 & $\beta^d_0 + \beta^d_1 x_{1n} + \beta^d_2 x_{2n}$ & -329.142 & 0.891 (0.003) & 0.889 (0.011) \\
M4 & $\beta^d_0 + \beta^d_1 x_{1n}$ & 277.355 & 0.567 (0.033) & 0.562 (0.039) \\
M5 & $\beta^d_0 + \beta^d_2 x_{2n}$ & 326.609 & 0.424 (0.046) & 0.420 (0.050) \\
M6 & $\beta^d_0 + \beta^d_4 x_{4n}$ & 430.729 & 0.011 (0.012) & 0.011 (0.012) \\
\midrule
\multicolumn{5}{c}{GAM} \\
\midrule
Code & Model & $WAIC$ & BR-CoDa-$R^2$ & BM-CoDa-$R^2$ \\
\midrule
M1 & $s^d(xs_{1n}, k = 10) + t^d(xs_{2n}, xs_{3n}, k = 10)$ & -690.713 & 0.989 (0.001) & 0.989 (0.002) \\
M2 & $s^d(xs_{1n}, k = 5) + t^d(xs_{2n}, xs_{3n}, k = 5)$ & -416.114 & 0.950 (0.004) & 0.948 (0.006) \\
M3 & $t^d(xs_{2n}, xs_{3n}, k = 10)$ & 87.851 & 0.264 (0.048) & 0.261 (0.050) \\
M4 & $t^d(xs_{2n}, xs_{3n}, k = 5)$ & 101.225 & 0.222 (0.048) & 0.219 (0.050) \\
M5 & $s^d(xs_{1n}, k = 10)$ & -99.864 & 0.779 (0.016) & 0.773 (0.022) \\
M6 & $s^d(xs_{1n}, k = 5)$ & -80.950 & 0.756 (0.018) & 0.750 (0.023) \\
\botrule
\end{tabular}
\end{table}
{In the linear CoDa regression setting, model M1, which includes all the terms in the data-generating process, shows the highest explained variability according to CoDa-$R^2$, in agreement with the WAIC results.} We computed the probabilities $P(\text{CoDa-}R^2_{M_1} \geq \text{CoDa-}R^2_{M_j})$ for $j = 2,\ldots,6$, using both BR-CoDa-$R^2$ and BM-CoDa-$R^2$. In all cases, these probabilities were equal to 1, indicating that M1 consistently explains more variability than any of the other models. In contrast, model M6, which included a covariate irrelevant to the data-generating process, shows the weakest performance. We also computed WAIC and the probabilities $P(\text{CoDa-}R^2_{M6} \geq \text{CoDa-}R^2_{M_j})$ for $j = 1,\ldots,5$, which were equal to 0 in all cases. {These results indicate that M6 explains substantially less variability than the other models.} Models M2 and M3 account for similar amounts of variability: the probabilities $P(\text{CoDa-}R^2_{M2} \geq \text{CoDa-}R^2_{M3})$ were 0.424 for BR-CoDa-$R^2$ and 0.477 for BM-CoDa-$R^2$, suggesting no relevant difference in the variability explained by both models.

{Additionally, the nested structure of the models allows us to examine how the inclusion of additional covariates changes the explained variability. For instance, in the linear regression example from \autoref{table:models_sim}, M3 is nested within both M1 and M2. Therefore, we can compare the corresponding reduced and extended models to quantify the increase in CoDa-$R^2$ associated with adding $x_3$ (in M1) or $x_4$ (in M2). However, when covariates are correlated, such incremental changes in explained variability should be interpreted with caution. The increase attributed to a given variable may depend on the set of variables already included in the model, and therefore on the order in which variables are added. As a result, these comparisons should be understood as conditional on the model specification rather than as an absolute measure of variable importance.}

{Adding $x_3$ to model M3 yields model M1. The probability $P(\text{CoDa-}R^2_{M1} \geq \text{CoDa-}R^2_{M3})$ was 1 for both BR-CoDa-$R^2$ and BM-CoDa-$R^2$, indicating that M1 explains substantially more variability than M3 under this model specification. On average, this corresponds to an increase of approximately 10\% in explained variability.}
In contrast, adding $x_4$ to M3 results in model M2. The probability $P(\text{CoDa-}R^2_{M2} \geq \text{CoDa-}R^2_{M3})$ is 0.424 for BR-CoDa-$R^2$ and 0.477 for BM-CoDa-$R^2$, indicating no relevant difference in explained variability. {Consequently, M2 and M3 provide a comparable level of explained variability according to CoDa-$R^2$, while WAIC slightly favors M3.}

In the GAM case, model M1, which includes both univariate and bivariate smooth terms with 10 basis functions for $xs_1$, $xs_2$, and $xs_3$, explains substantially more variability than the other models, as the $P(\text{CoDa-}R^2_{M1} \geq \text{CoDa-}R^2_{M_j})$, $j=2,\ldots,6$, is 1. This is consistent with the WAIC results. On the other hand, models including only the bivariate smooth terms (M3 and M4) explain the least variability. The resulting probabilities, $P(\text{CoDa-}R^2_{M3} \geq \text{CoDa-}R^2_{M4})$, are 0.730 for BR-CoDa-$R^2$ and 0.707 for BM-CoDa-$R^2$. Thus, we conclude that both models explain a similar amount of variability, and that the impact of covariates $xs_2$ and $xs_3$ on the response variable is lower than that of $xs_1$. As in the linear CoDa regression case, the nested structure of models M3 and M5 within M1, and M4 and M6 within M2, allows us to quantify the variability explained by each term when transitioning from M3 or M5 to M1, or from M4 or M6 to M2. Lastly, in line with the univariate proposal by Vehtari et al. \cite{vehtariBayesR2}, we also observed that the distribution of BR-CoDa-$R^2$ remains narrower than that of BM-CoDa-$R^2$. {Overall, these results illustrate how CoDa-$R^2$ can be used to summarize differences in explained variability across models, while WAIC provides the main predictive criterion for assessing model adequacy.}

\section{A Bayesian CoDa-GAM for soil texture prediction in the Basque Country}
\label{sec:CoDAGAMSoil}

Let $\boldsymbol{y}_{n}$ be a 3-dimensional compositional vector representing the proportions of sand ($y_{1n}$), silt ($y_{2n}$), and clay ($y_{3n}$). The model under study is then defined as
\begin{equation}
\label{eq:model_soil}
\begin{split}
\boldsymbol{y}^*_{n} &= \mathrm{ilr}(\boldsymbol{y}_{n}) \sim \mathcal{N}(\boldsymbol{\mu}_{n}, \boldsymbol{\Sigma}), \\
\mu_{dn} &= \beta_0^d + \beta_1^d \operatorname{Lit}_n + \beta_2^d \operatorname{Year}_n + s_1^d(\operatorname{Elev}_n) + s_2^d(\operatorname{Slope}_n) + t^d(\operatorname{Lon}_n, \operatorname{Lat}_n),
\end{split}
\end{equation}
where $\boldsymbol{y}^*_{n} = [y_{1n}^*, y_{2n}^*]$ is the 2-dimensional ilr coordinate vector of the $n$th observation. The ilr coordinates correspond to the following balance-based representation (see \autoref{eq:ilr}):
\response{
\begin{equation}
\begin{split}
y_{1n}^* &= -\sqrt{\frac{1}{2}} \log \left( \frac{y_{1n}}{y_{2n}} \right) = \sqrt{\frac{1}{2}} \log \left( \frac{y_{2n}}{y_{1n}} \right), \\
y_{2n}^* &= -\sqrt{\frac{2}{3}} \log \left( \frac{(y_{1n}y_{2n})^{1/2}}{y_{3n}} \right) = \sqrt{\frac{2}{3}} \log \left( \frac{y_{3n}}{(y_{1n}y_{2n})^{1/2}} \right),
\end{split}
\end{equation}
}
We note that the first ilr coordinate quantifies the relative proportion of silt to sand, whereas the second ilr coordinate measures the balance between the geometric mean of sand and silt relative to clay. Lit$_n$ (Lithology) is modeled as a fixed effect, Year$_n$ (Year) as a random effect, $s^d(\bullet)$ denotes a univariate penalized P-spline, and $t^d(\bullet)$ represents a bivariate tensor product in a penalized P-spline basis. {For all smooth terms, we used a basis dimension of $k = 10$, chosen to be sufficiently large to allow flexible fits. In practice, model complexity is primarily controlled through the smoothing penalty, so this choice ensures that the basis dimension does not constrain the estimation of nonlinear effects.} 

\response{Priors for the fixed-effect parameters, including the linear covariates and the unpenalized (linear) component \(\boldsymbol{\beta}_j^F\) of each P-spline term, were set to Gaussian distributions centered at 0 with a standard deviation of 10. The variance components \(\sigma_j^2\) governing the penalized part of each smooth term were assigned \texttt{brms}'s default half-Student-\(t\) prior with three degrees of freedom, consistent with the general specification described above. The same default prior was used for the remaining standard deviations, and the correlation matrix was assigned an LKJ prior with regularization parameter equal to 1.} {Model comparison was based primarily on WAIC, while the proposed Bayesian CoDa-$R^2$ measures were used to summarize the explained variability and complement the model assessment.}

\subsection{{Model evaluation}}

We compared several candidate models using WAIC, BR-CoDa-$R^2$, and BM-CoDa-$R^2$ (see \autoref{table:models_soil} and \autoref{fig:r-squared_texture}). {Model M1, which includes spatial effects (longitude and latitude), Lithology, Elevation, Slope, and a random Year effect, achieved the lowest WAIC value and also showed the highest explained variability according to both BR-CoDa-$R^2$ and BM-CoDa-$R^2$.} 

\response{The posterior probabilities $P(\text{CoDa-}R^2_{M1} \geq \text{CoDa-}R^2_{Mj})$, $j = 2,\ldots,6$, were
0.969, 1, 1, 1, and 1 for BR-CoDa-$R^2$, and 0.957, 1, 0.997, 1, and 1 for
BM-CoDa-$R^2$, indicating that M1 consistently showed higher explained
variability than the alternatives.} Comparing M1 and M2 shows that including the random Year effect is associated with an average increase of approximately $3\%$ in explained variability under both BR-CoDa-$R^2$ and BM-CoDa-$R^2$. This highlights the relevance of accounting for inter-annual variability in soil texture modeling. \response{Similarly, comparing M1 and M3 shows that including the spatial term is
associated with a much larger increase in explained variability, of
approximately $14\%$ under both BR-CoDa-$R^2$ and BM-CoDa-$R^2$,
indicating that, under the pragmatic decision rule introduced in Section 6, the spatial term is the most influential component among those compared here.} {\autoref{fig:r-squared_texture} illustrates that M1 achieved the highest mean CoDa-$R^2$, in agreement with the WAIC-based comparison.}

\begin{table}[h]
\caption{Summary of the models fitted to the soil texture dataset. For each model, the WAIC and the posterior mean (with standard deviation) of the BR-CoDa-$R^2$ and BM-CoDa-$R^2$ are reported.}
\label{table:models_soil}
\centering
\begin{tabular}{lcccccccc}
\hline
Code & Lit & Elev & Slope & Lon, Lat & Year & WAIC & BR-CoDa-$R^2$ & BM-CoDa-$R^2$ \\
\hline
M1 & $\checkmark$ & $\checkmark$ & $\checkmark$ & $\checkmark$ & $\checkmark$ & 2369.722 & 0.346 (0.011) & 0.346 (0.012) \\
M2 & $\checkmark$ & $\checkmark$ & $\checkmark$ & $\checkmark$ & $-$ & 2607.442 & 0.315 (0.011) & 0.315 (0.013) \\
\response{M3} & \response{$\checkmark$} & \response{$\checkmark$} & \response{$\checkmark$} & \response{$-$} & \response{$\checkmark$} & \response{3227.734} & \response{0.209 (0.011)} & \response{0.209 (0.012)} \\
M4 & $-$ & $\checkmark$ & $\checkmark$ & $\checkmark$ & $-$ & 2715.712 & 0.299 (0.012) & 0.298 (0.013) \\
M5 & $-$ & $-$ & $-$ & $\checkmark$ & $-$ & 2773.772 & 0.284 (0.011) & 0.284 (0.012) \\
M6 & $\checkmark$ & $-$ & $-$ & $-$ & $-$ & 3986.285 & 0.071 (0.007) & 0.071 (0.008) \\
\hline
\end{tabular}
\end{table}
\begin{figure}[h]
\centering
\includegraphics[width=\textwidth]{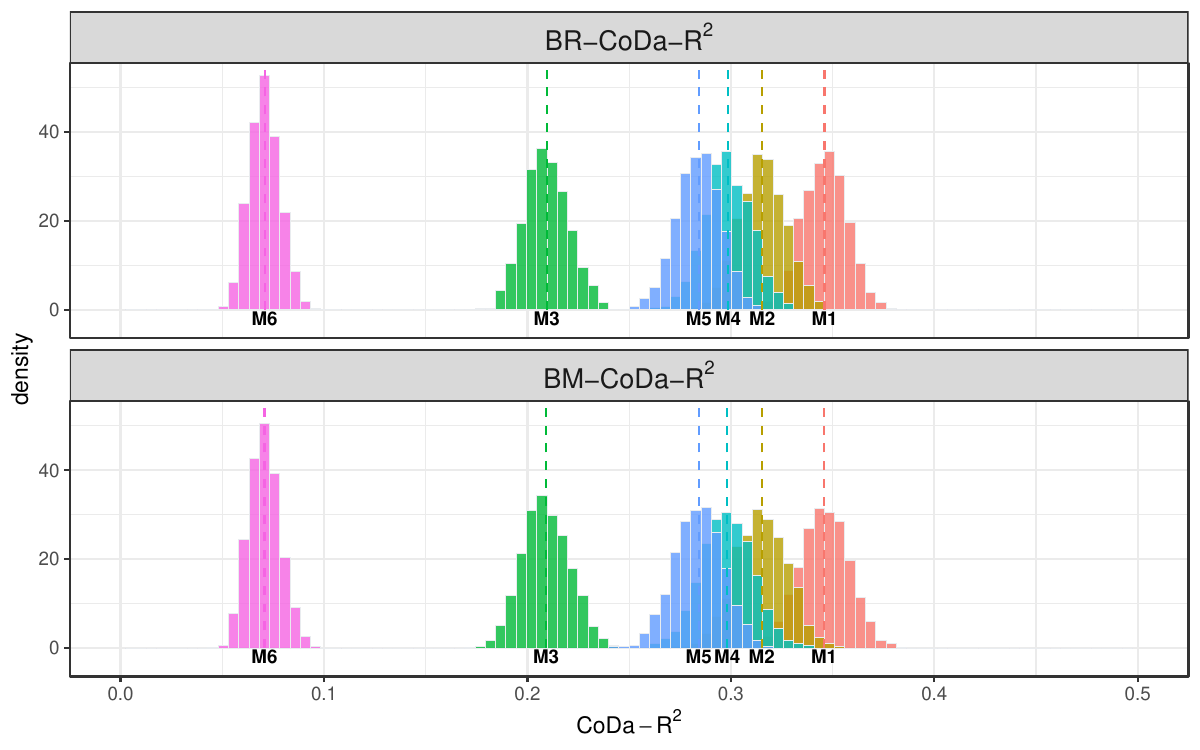}
\caption{Posterior distributions of BR-CoDa-$R^2$ and BM-CoDa-$R^2$ for each of the soil texture models.}
\label{fig:r-squared_texture}
\end{figure}
\response{Building on the selection of M1, we further examined whether the smooth term for Slope could be simplified to a linear effect. The two specifications yielded WAIC values of 2369.722 (smooth Slope) and 2364.316 (linear Slope), and comparable explained variability: BR-CoDa-$R^2$ of 0.346 (0.011) versus 0.345 (0.011), and BM-CoDa-$R^2$ of 0.346 (0.012) versus 0.345 (0.012). The corresponding posterior probabilities that the linear specification attains a higher CoDa-$R^2$ than the smooth one were 0.502 (BR-CoDa-$R^2$) and 0.498 (BM-CoDa-$R^2$), both well within the $(0.1, 0.9)$ range indicating comparable explanatory capacity under the decision rule introduced in \autoref{sec:BayesianRsquared}. Given this equivalence, and the added interpretability of a directly estimated coefficient, we adopt the linear specification for Slope in the model reported below.}

\subsection{{Selected model and predictions}}

After selecting the final model to represent the process of interest, we computed the posterior distribution of the fixed effects and smooth terms. The \texttt{brms} package allowed us to obtain these in a straightforward way. Estimates were obtained in real space, the domain where inference is carried out. While fixed effects retain a direct interpretation in this space, smooth terms are more naturally understood in the original compositional space $\mathcal{S}^3$. Owing to the isometric property of the ilr transformation, it suffices to apply the inverse ilr transformation directly to the posterior simulations to recover results in the simplex.

In \autoref{table:litho_fixed}, we report the posterior distributions of the parameters corresponding to the fixed effect of Lithology. Category 1 (Surface deposits) serves as the reference level. For the first ilr coordinate, which quantifies the log-ratio between silt and sand, the largest positive contribution is associated with category 6 (Decarbonate marls). The posterior mean of the parameter indicates that, for decarbonate marls, the \response{silt-to-sand ratio is on average 54\% higher than in surface deposits}.
For the second ilr coordinate, which measures the balance between the geometric mean of sand and silt relative to clay, category 12 (Limestones and dolomites) shows the highest positive contribution.  The posterior mean suggests that, under this lithology, the \response{clay proportion relative to the combined (geometric-mean) proportion of sand and silt is approximately double that in surface deposits}.

\response{For the Slope covariate, now included as a linear fixed effect, both posterior coefficients are positive, indicating that steeper terrain is associated with a higher silt-to-sand ratio and a higher clay proportion relative to the combined sand and silt proportion. Each one-percentage-point increase in slope is associated with an average 0.14\% increase in the silt-to-sand ratio and a 0.37\% increase in the clay-to-(sand, silt) ratio.}
\begin{table}[h]
\caption{Posterior distribution of the parameters corresponding to the fixed effect Lithology and Slope. Category 1 (Surface deposits) serves as the reference level.}
\label{table:litho_fixed}
\centering
\response{
\begin{tabular}{lrrrrrr}
\hline
 & \multicolumn{3}{c}{ilr1} & \multicolumn{3}{c}{ilr2} \\
\hline
Lithology & Mean & Q 2.5\% & Q 97.5\% & Mean & Q 2.5\% & Q 97.5\% \\
\hline
Lito2 & 0.006 & -0.128 & 0.136 & 0.067 & -0.111 & 0.250 \\
Lito3 & 0.146 & -0.098 & 0.384 & -0.009 & -0.342 & 0.299 \\
Lito4 & 0.033 & -0.123 & 0.187 & 0.122 & -0.086 & 0.324 \\
Lito5 & 0.043 & -0.019 & 0.102 & -0.054 & -0.137 & 0.029 \\
Lito6 & 0.304 & 0.093 & 0.505 & 0.180 & -0.101 & 0.457 \\
Lito7 & 0.158 & 0.091 & 0.226 & 0.203 & 0.113 & 0.292 \\
Lito8 & 0.152 & 0.098 & 0.205 & 0.126 & 0.054 & 0.199 \\
Lito9 & -0.049 & -0.127 & 0.024 & 0.238 & 0.132 & 0.343 \\
Lito10 & 0.156 & 0.116 & 0.197 & 0.136 & 0.080 & 0.189 \\
Lito11 & 0.021 & -0.178 & 0.226 & -0.006 & -0.277 & 0.246 \\
Lito12 & 0.010 & -0.308 & 0.319 & 0.575 & 0.147 & 1.012 \\
Lito13 & -0.046 & -0.174 & 0.090 & 0.158 & -0.032 & 0.341 \\
Slope & 0.001 & 0.000 & 0.002 & 0.003 & 0.002 & 0.005 \\ 
\hline
\end{tabular}
}
\end{table}

\response{\autoref{fig:smoothterms_elev} displays the posterior marginal distribution of the smooth Elevation effect on the ilr-transformed soil texture coordinates and their corresponding compositional proportions. The plots reveal that higher elevations correspond to a slight increase in sand, a more pronounced increase in silt, and a decrease in clay.} Thanks to the Bayesian framework, we can quantify the associated uncertainty. \response{Uncertainty in the elevation effect is reduced around 500 meters.} {From an environmental perspective, these patterns are consistent with known soil formation processes. Areas with higher clay content, particularly those associated with steeper slopes, tend to exhibit greater water retention but also increased susceptibility to compaction and reduced drainage capacity. This has direct implications for agricultural management, as such soils may require specific practices to prevent waterlogging and maintain soil structure. In contrast, areas with higher sand proportions are typically associated with better drainage but lower nutrient retention, which can limit crop productivity without appropriate soil amendments.}

\begin{figure}[h]
\centering
\includegraphics[width=\textwidth]{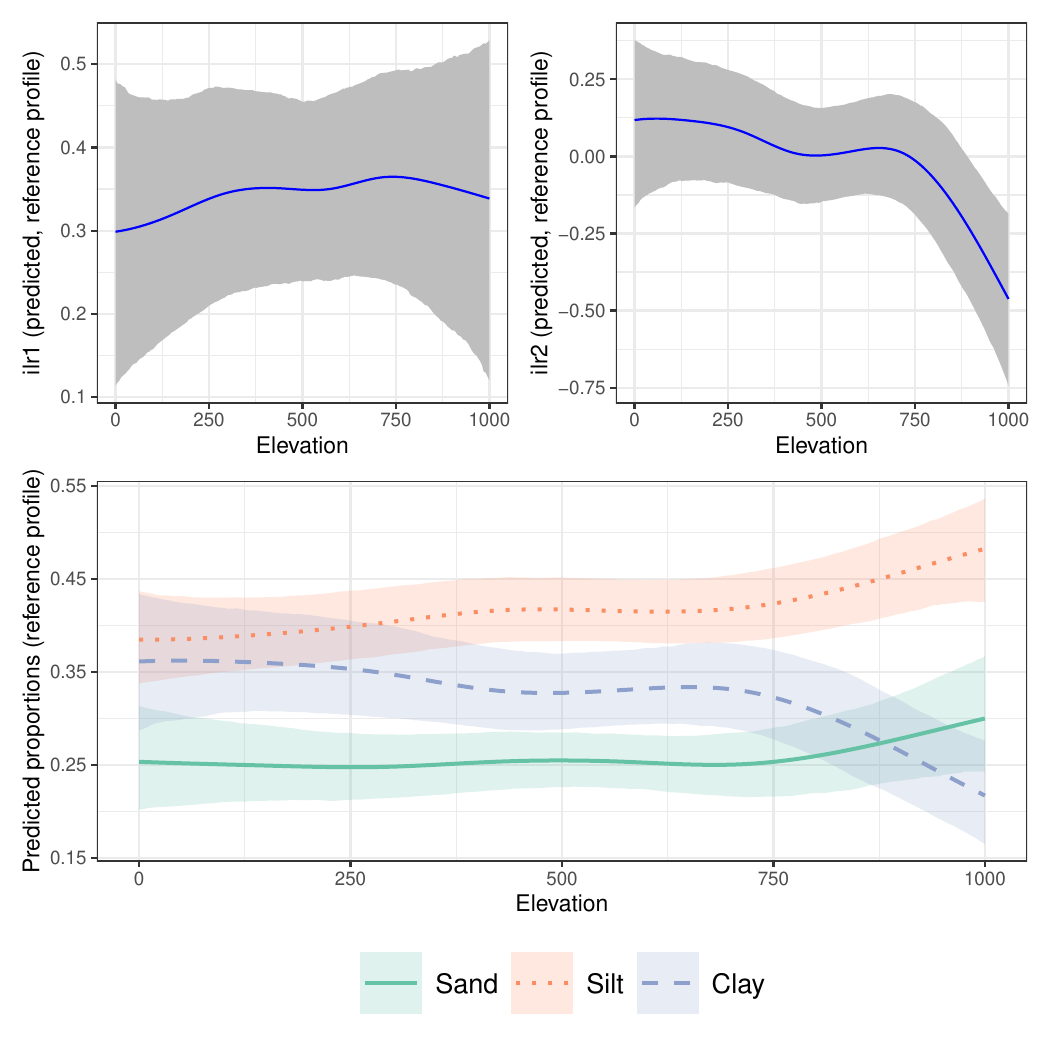}
\caption{Marginal posterior distribution of the smooth effect of elevation on the ilr-transformed soil texture coordinates and their transformation to the compositional space.}
\label{fig:smoothterms_elev}
\end{figure}
The posterior distributions of the two-dimensional smooth spatial terms are shown in \autoref{fig:smoothterms_spatial}. These plots reveal clear spatial patterns in the composition of sand, silt, and clay across the study region. For sand, higher proportions are predicted in specific geographic areas, likely corresponding to zones of lower elevation or particular lithologies. In contrast, silt shows increased proportions in areas where sand decreases, reflecting the natural balance in CoDa. Clay proportions appear more homogeneous across space, with no strong spatial gradient detected. {These spatial patterns can be directly interpreted in terms of land use and environmental processes. For instance, regions with higher sand proportions may be more prone to water stress, while areas dominated by clay may be at higher risk of compaction or reduced infiltration. Identifying these spatial gradients is particularly relevant for precision agriculture and environmental planning, as it allows targeting specific interventions based on local soil composition.}

{A key advantage of the Bayesian framework is the ability to explicitly quantify predictive uncertainty and incorporate it into decision-making. Regions with higher posterior uncertainty should be interpreted with caution, particularly when predictions are used for environmental risk assessment or management decisions. In contrast, areas with low uncertainty provide more reliable information for planning purposes, such as identifying suitable zones for agricultural use or assessing vulnerability to soil degradation processes.}

\begin{figure}[h]
\centering
\includegraphics[width=\textwidth]{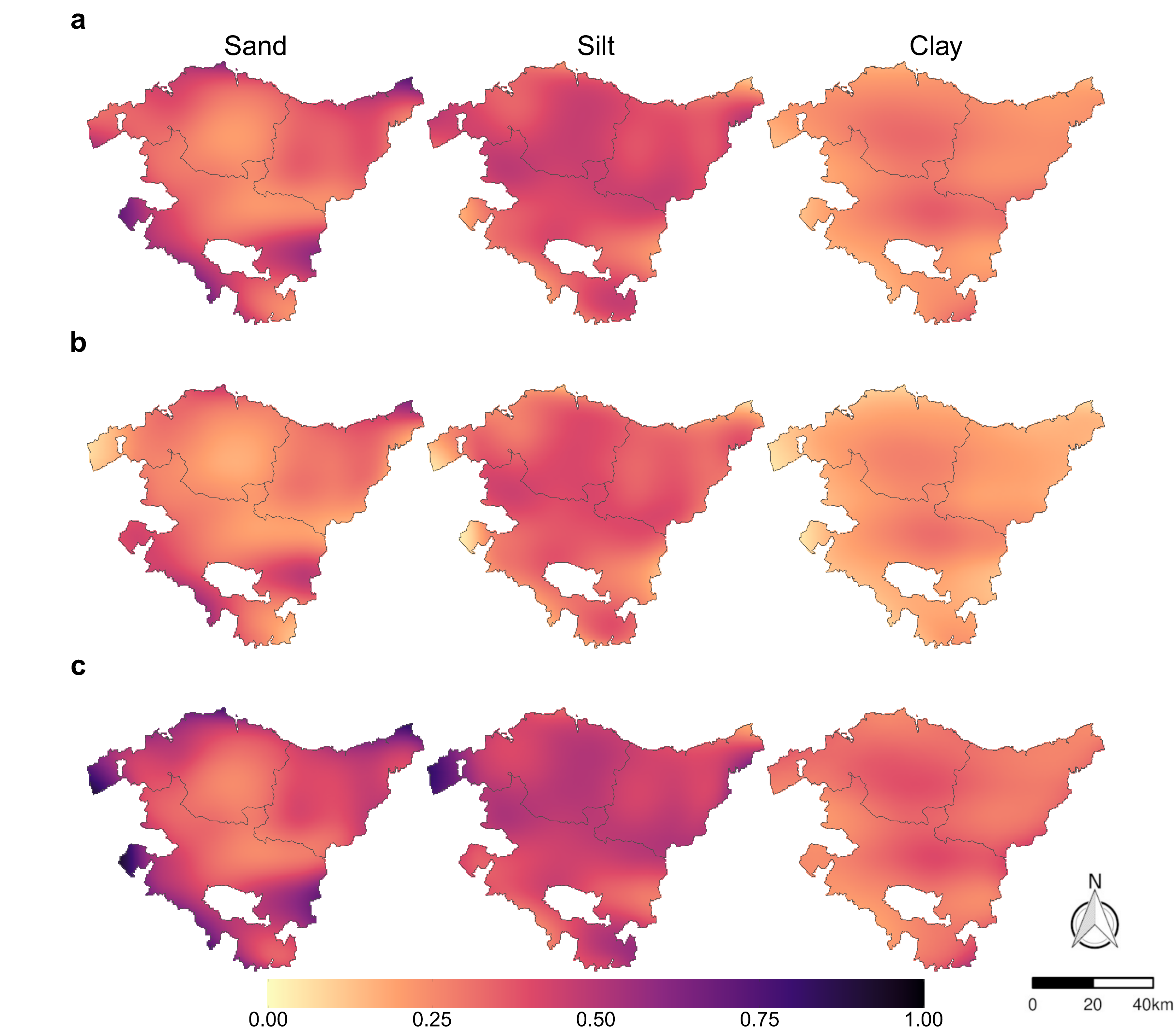}
\caption{Posterior distribution of the smooth spatial term in the original space. Mean and 95\% credible intervals are depicted.}
\label{fig:smoothterms_spatial}
\end{figure}

Finally, we produced predictions for the entire Basque Country, including the mean and standard deviation of the posterior predictive distributions, as well as the USDA textural classification derived from them (\autoref{fig:prediction_texture}). A key advantage of our modeling framework is the straightforward computation of predictive uncertainty. This stems from the isometric nature of the ilr transformation, which preserves distances and variances when mapping between the simplex and real space. Therefore, applying the inverse ilr transformation to posterior predictive samples in real space directly yields the full predictive distribution in the simplex. This enables interpretation in terms of soil texture proportions and supports classification according to the USDA textural system, a widely adopted standard in soil science that categorizes soils by their relative contents of sand, silt, and clay \citep{united1993soil}. {In our implementation, the USDA textural class at each location is derived from the posterior mean predictive composition, providing a point estimate that is easy to interpret and communicate for practical applications. While uncertainty is not directly propagated to the USDA class itself, the Bayesian framework allows uncertainty to be quantified for the predicted soil composition, which is valuable for land suitability assessment and related environmental applications.} In this study, we computed the USDA textural class for each prediction point, generating a comprehensive soil texture classification map across the Basque Country (\autoref{fig:prediction_texture}). {Compared to standard approaches that model each soil component independently, such as separate kriging or regression models, the proposed framework ensures that predictions remain coherent compositions while jointly propagating uncertainty across components, avoiding inconsistencies that may arise when components are modeled in isolation.}

\begin{figure}[H]
\centering
\includegraphics[width=\textwidth]{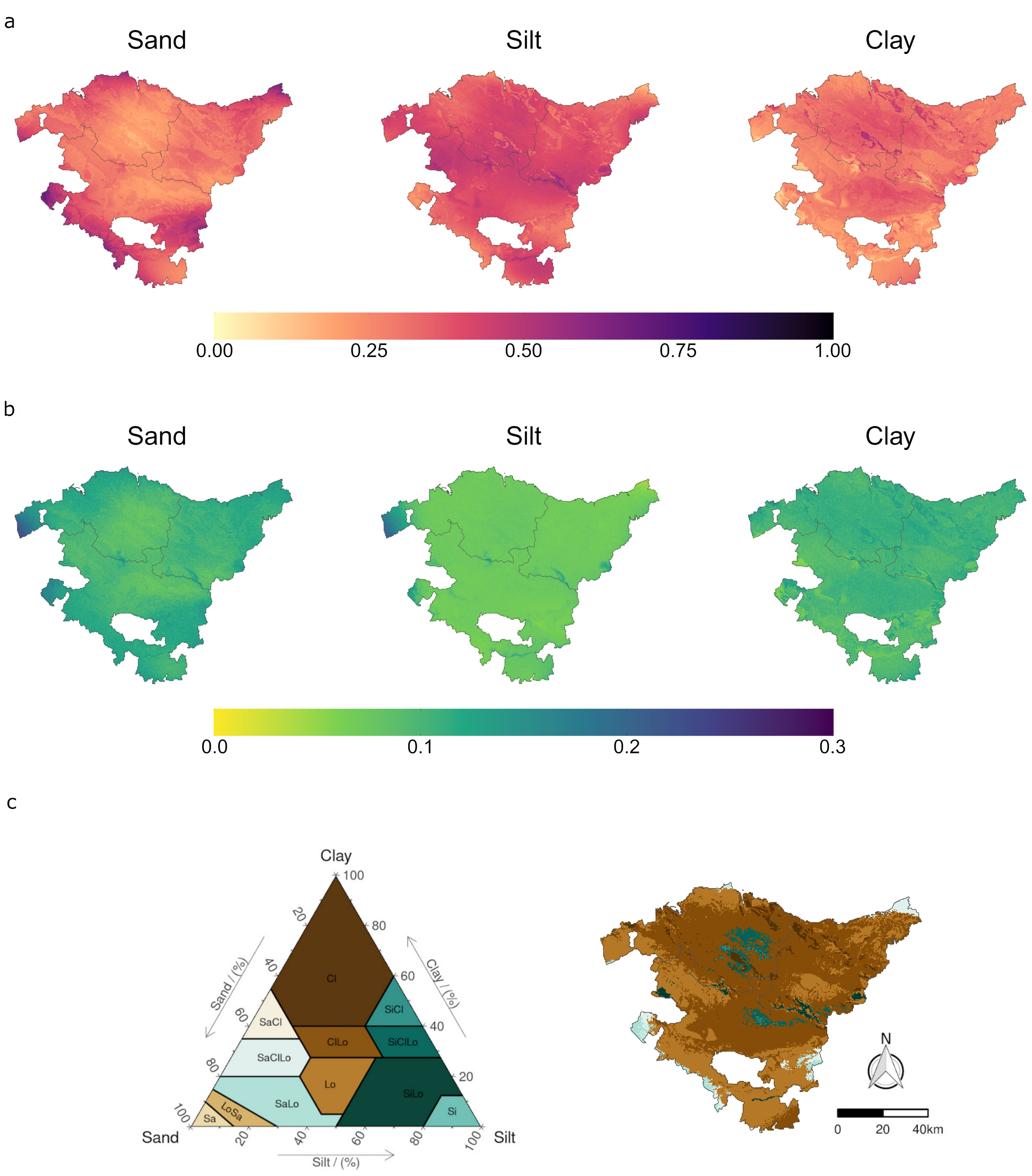}
\caption{(a) Mean posterior predictive distribution for sand, silt and clay proportions. (b) Standard deviations of the posterior predictive distribution. (c) USDA textural classification.}
\label{fig:prediction_texture}
\end{figure}

\section{Concluding remarks}\label{sec:Conclusions}

This work addressed key challenges in the modeling of spatial CoDa by proposing Bayesian GAM models for CoDa using penalized splines and tensor products within a flexible regression framework. Two main contributions were presented. First, we developed a Bayesian modeling framework that leverages the ilr transformation to incorporate smooth terms and spatial effects while preserving the geometry of the simplex. {These measures are intended as interpretable in-sample summaries of explained variability and are best used alongside predictive criteria such as WAIC.} 

Simulation studies demonstrated the capacity of the proposed models to accurately recover underlying parameters and quantify explained variability. The real data application, predicting soil texture composition in the Basque Country, showcased the practical relevance of these models, highlighting the importance of spatial effects and covariates such as lithology, elevation, and slope. The \texttt{brms} package, together with our CoDa-specific extensions, facilitated both implementation and interpretation of results in the simplex space. {The proposed framework yields more coherent and interpretable predictions than approaches that model each component independently. It enables the identification of relevant geographical patterns for land use planning and soil management. Moreover, the Bayesian formulation enables a full characterization of predictive uncertainty, which is essential for risk assessment and decision-making in environmental contexts.}

{While these results highlight the practical usefulness of the proposed framework, some methodological considerations should be noted. In particular, although the proposed Bayesian geoadditive modeling strategy can employ alternative log-ratio transformations (e.g., alr or clr), the CoDa-$R^2$ measures presented here critically rely on the isometric property of the ilr transformation. In addition, since different ilr representations are related through orthogonal transformations, the proposed CoDa-$R^2$ measures are invariant to the particular choice of ilr basis, including the ordering of the compositional parts.} Unlike ilr, both alr and clr are not isometries: they distort distances and angles relative to the Aitchison geometry, preventing a coherent decomposition of variability and rendering any derived $R^2$ values difficult to interpret in terms of the original compositional structure. Therefore, the CoDa-$R^2$ measures proposed in this work are not directly transferable to models using alr or clr transformations.

The proposed BM-CoDa-$R^2$ relies on the assumption that the ilr-transformed coordinates of the response variable follow a multivariate normal distribution, as is standard in Gaussian Bayesian modeling. In contrast, the BR-CoDa-$R^2$ only requires posterior predictive samples of the ilr coordinates, and can therefore be used as a general-purpose validation metric in any Bayesian predictive framework that provides such samples. This includes flexible Bayesian machine learning methods such as Bayesian additive regression trees (BART), Bayesian neural networks, or other simulation-based Bayesian approaches, provided that posterior predictive distributions in the real (ilr) space are available. Exploring these possibilities is a promising avenue for future research, especially in settings where classical parametric assumptions may not hold.

\section*{Declarations}

\begin{itemize}
\item \textbf{Funding} Joaquín Martínez-Minaya, Lore Zumeta-Olaskoaga and Dae-Jin Lee gratefully acknowledge the Ministry of Science, Innovation and Universities (Spain) for research project PID2020-115882RB-I00.

\item \textbf{Conflict of interest/Competing interests} Not applicable.

\item \textbf{Ethics approval and consent to participate} Not applicable.

\item \textbf{Consent for publication} Not applicable.

\item \textbf{Data availability} Not applicable.

\item \textbf{Materials availability} \response{The supplementary sensitivity analyses referred to throughout this manuscript are provided as Online Resource 1 (noise and correlation sensitivity) and Online Resource 2 (prior sensitivity), submitted alongside this article.}

\item \textbf{Code availability} All code used to implement the models described in the simulation studies is publicly available at the following GitHub repository: \url{https://github.com/jmartinez-minaya/CoDabrms/}

\item \textbf{Author contribution} Not applicable.
\end{itemize}

\begin{appendices}

\section{Additional Figures and Tables}\label{appn}

\begin{figure}[H]
    \centering
    \includegraphics[width=\textwidth]{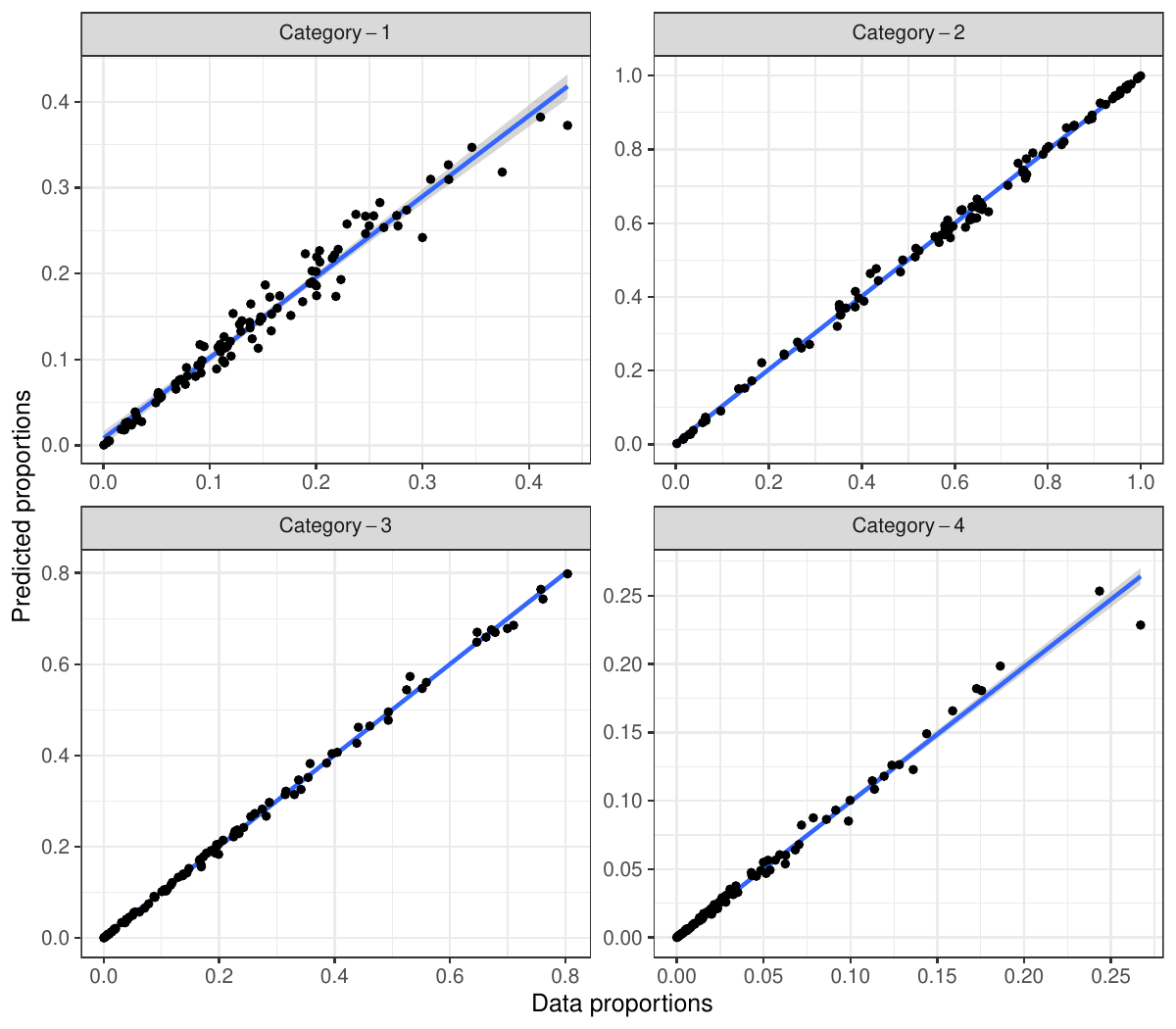}
    \caption{Posterior predictive checks for the linear regression simulation example. Real values \textit{vs.} mean of the posterior predictive distribution for each category in the composition. The blue line corresponds to a linear regression fit; proximity to the identity line indicating the model's ability to capture the linear structure.}
    \label{fig:fitted_linear_regression}
\end{figure}

\begin{figure}[H]
    \centering
    \includegraphics[width=\textwidth]{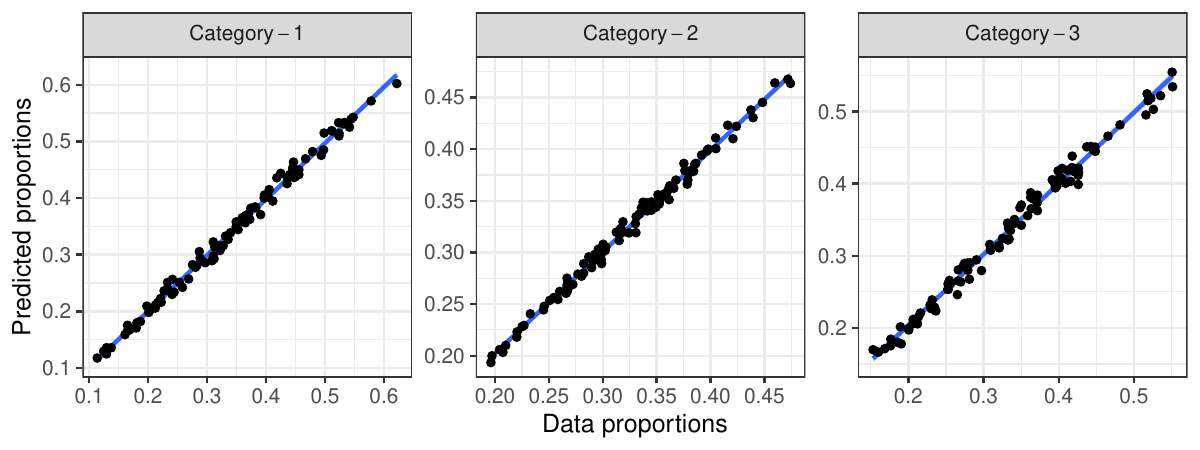}
    \caption{Posterior predictive checks for the GAM simulation example. Real values \textit{vs.} mean of the posterior predictive distribution for each compositional category. The blue line represents the linear regression fit between observed and predicted values, illustrating the model's ability to capture the nonlinear structure.}
    \label{fig:fitted_sim_splines}
\end{figure}

\end{appendices}

\bibliography{compositional}

\end{document}